\documentclass[aps,prb,twocolumn,superscriptaddress,unsortedaddress]{revtex4-2}

\usepackage{graphicx}
\usepackage{dcolumn}
\usepackage{bm}
\usepackage[tight]{subfigure}
\usepackage{enumerate}
\usepackage{amsmath}
\usepackage{verbatim}
\usepackage{color}
\usepackage{natbib}
\usepackage{float}
\usepackage{siunitx}

\begin{document}


\title{Spectroscopy of the local density-of-states in nanowires using integrated quantum dots}

\author{Frederick S. Thomas}
\author{Malin Nilsson}
\author{Carlo Ciaccia}
\author{Christian J{\"u}nger}
\affiliation{Department of Physics, University of Basel, Klingelbergstrasse 82, CH-4056 Basel, Switzerland}

\author{Francesca Rossi}
\affiliation{IMEM-CNR, Parco Area delle Scienze 37/A, I-43124 Parma, Italy}

\author{Valentina Zannier}
\author{Lucia Sorba}
\affiliation{NEST, Istituto Nanoscienze-CNR and Scuola Normale Superiore, Piazza San Silvestro 12, I-56127 Pisa, Italy}

\author{Andreas Baumgartner}
\author{Christian Sch{\"o}nenberger} \email{Christian.Schoenenberger@unibas.ch} \homepage[]{www.nanoelectronics.ch}
\affiliation{Department of Physics, University of Basel, Klingelbergstrasse 82, CH-4056 Basel, Switzerland}
\affiliation{Swiss Nanoscience Institute, University of Basel,Klingelbergstrasse 82, CH-4056 Basel, Switzerland}




\date{\today}

\begin{abstract}
In quantum dot (QD) electron transport experiments additional features can appear in the differential conductance $dI/dV$ that do not originate from discrete states in the QD, but rather from a modulation of the density-of-states (DOS) in the leads. These features are particularly pronounced when the leads are strongly confined low dimensional systems, such as in a nanowire (NW) where transport is one-dimensional and quasi-zero dimensional lead-states can emerge.
In this paper we study such lead-states in InAs NWs. We use a QD integrated directly into the NW during the epitaxial growth as an energetically and spatially well-defined tunnel probe to perform $dI/dV$ spectroscopy of discrete bound states in the `left' and `right' NW lead segments. By tuning a side\-gate in close proximity of one lead segment, we can distinguish transport features related to the modulation in the lead DOS and to excited states in the QD.
We implement a non-interacting capacitance model and derive expressions for the slopes of QD and lead resonances that appear in two-dimensional plots of $dI/dV$ as a function of source-drain bias and gate voltage in terms of the different lever arms determined by the capacitive couplings. We discuss how the interplay between the lever arms affect the slopes. We verify our model by numerically calculating the $dI/dV$ using a resonant tunneling model with three non-interacting quantum dots in series. Finally, we used the model to describe the measured $dI/dV$ spectra and extract quantitatively the tunnel couplings of the lead segments. Our results constitute an important step towards a quantitative understanding of normal and superconducting subgap states in hybrid NW devices.
\end{abstract}


\maketitle

\section{Introduction}

Semiconducting nanowires (NWs) coupled to superconductors~\cite{DeFranceschi2010} are intensively being explored as a system for the realization of Majorana fermions (MFs) which acts as building blocks for topological protected quantum computing.~\cite{Stanescu2013,Lutchyn2018}
Here, part of the NW is covered by a thin film of a conventional s-wave superconductor. The interplay between an external magnetic field, spin-orbit interaction in the semiconductor, and pairing interaction gives rise to topological superconductivity in a certain parameter window. Majorana bound state (MBSs) are then expected to appear at the two ends of the superconducting region. Multiple approaches have been proposed to perform quantum state operations and read-out of MBSs.~\cite{Sau2010,Beenakker2013, Hyart2013,Aasen2016,Plugge2017}
Among them are quantum dots (QDs) that act as an element for a controlled coherent coupling of the fermion degree of freedom in the QD to a MBS. They can be used to couple, for example, different NW segments or for the readout of Majorana-based qubit states.~\cite{Leijnse2011,Liu2011,Deng2016,Aasen2016,Plugge2017}
In most experiments with NWs, a tunnel barrier has been defined close to the end of the superconducting region by metallic (side-) gates that `locally' deplete the NW. In a charge transport experiment, with charge current $I$, most of the applied source-drain bias voltage $V_{sd}$ is then supposed to develop across the tunnel junction between the normal lead and the superconducting one. This allows one to perform tunneling spectroscopy where the differential conductance $dI/dV$ is measured as a function of the source-drain voltage $V_{sd}$, which serves as a tool to study proximity-induced superconductivity and the evolution of sub-gap states into MBSs. However, due to a finite diameter and finite resolution in lithography, there is always a NW segment in between the tunnel barrier and the region of the NW covered with the superconductor. Since the induced potential landscape depends on many factors, like the NW diameter, the gate configuration, the different work-functions of the evaporated metals and contacts, there always remains an ambiguity in the location of the tunnel barrier and the location of a possible MBS, which on its own extends into the normal and superconducting region with some characteristic decay length. The softer the potential, the bigger this ambiguity. This is the reason why spurious QDs often appear in NW experiments, in particular, if the NW is used close to pinch-off at low carrier concentration. Recently, the research community started to appreciate and accept that there is a potential well in between the tunnel barrier and the proximitized NW, which can host a single or multiple Andreev bounds states (ABSs). In experiments, these ABSs may evolve into zero-bias anomalies with an increasing external magnetic field.~\cite{Liu2016}


Unlike gate-defined tunnel barriers and QDs, epitaxially defined barriers that are integrated during the growth of the NWs can be very sharp.~\cite{Seifert2004,Bjork2002} In epitaxial growth the material can be switched from, for example, InAs to InP within one monolayer in the best case. For thin segments of InP, there is a large band-offset of $\sim\,0.4$\,eV in the conduction band between InAs and InP.~\cite{Niquet2008, Thomas2020} This results in a very strong confinement that cannot be realized with gate electrodes. Integrated QDs are therefore an ideal building block for spectroscopy of the emergence of proximity-induced superconductivity and to study the evolution of subgap states in InAs NWs.

Recently, epitaxially defined QDs, where the confinement is obtained by two NW segments in the wurtzite phase in otherwise zinc blende InAs NW,~\cite{Nilsson2016} were employed for tunnel spectroscopy of the local density of states in proximitized NWs.~\cite{Junger2019, Junger2020}
In these experiments, one can see that the proximity-induced gap appears sharply with gate voltage at an electron density for which the superconducting coherence length becomes larger than the distance between the integrated QD and the region covered by the superconductor. We will refer to this nanowire section in the following as the `lead segment'. Quasi-bound states can form in the lead segment that may turn into Andreev-bound or Yu-–Shiba-–Rusinov sugap-states in the superconducting case.\cite{Yu1965,Shiba1968,Rusinov1969}
It is known that quasi-bound states in the leads can result in conductance features in NW QD devices. In general, one can interpret these features as a modulation of the density-of-states (DOS) in the leads.~\cite{Bjork2003,Bjork2004,Bjork2005}
Thomas \textit{et al}. recently presented an in-depth study of the electrical characteristics of a QD defined by two InP segments in an InAs NW in the wurtzite phase.~\cite{Thomas2020} The QD could be loaded in a controlled way starting from pinch-off up to a few hundred of electrons. Although the potential step in the conduction-band onset, given by the band-offset between InAs and InP, does not change, the tunnel coupling between the NW lead and the QD was shown to be gate voltage dependent. This is caused by the increase of the Fermi energy with carrier density which lowers the effective tunneling barrier height. Close to pinch-off a life-time broadening $\Gamma$ of the QD state of only $0.5$\,$\mu$eV could be confirmed. For larger fillings $\Gamma$ reached values $> 600$\,$\mu$eV. The tuning of the tunnel coupling to the QD is an important aspect since it sets the  resolution of the spectroscopy.

In another work on similar devices, even lower tunnel coupling strength were reported reaching $120$\,$nV$ for the first electron. Interestingly, two qualitatively different regimes were found at low filling numbers and seen in a step-like transition of the peak conductance depending on the axial quantum number of orbital wavefunction.~\cite{Momtaz2020}

Efforts have been made to distinguish and study conductance features originating from lead segments in electron accumulation layers near a Si-SiO$_2$ interface using gate-defined QDs or a phosphorus donor ion as a tunnel spectroscopy probe~\cite{Mottonen2010} and in graphene leads where a carbon-based molecule acts as spectroscopy probe.~\cite{Gehring2017}
However, an in-depth electrostatic description of the origin and characteristics of those features has not yet been presented for NW devices. The epitaxially defined integrated QD provides a robust and tunable tool for a systematic tunnel spectroscopy of the local DOS in the NW lead segment.

In this paper, we use an integrated QD defined during growth by an axial InAs/InP heterostructure as a tunnel probe to study the states formed in the NW segment adjacent to the QD. Using two separate gates allows us to tune the chemical potential in the NW leads with different strengths, which provides a route to distinguish the lead resonances from excited states in the QD. We use a simple capacitance model to derive explicit expressions for the slopes of the resonances in the differential conductance $dI/dV$ and discuss the relation between the lever arms and the slopes of the resonances. We show that the main lead resonances appear when a quasi-bound state in either the source or drain lead segment aligns with a state in the QD. The combined effects of two resonances results in pronounced negative differential conductance. This is confirmed in a simple calculation of the differential conductance using a resonant tunneling model with the lever arms obtained from the experimental data as input. With this model we can simulate $dI/dV$ spectra that are in excellent agreement with the measurements.

\section{Experiment}

For the tunnel spectroscopy measurements, we use InAs/InP hetero\-structure NW with a diameter of $50\pm 5$\,nm.~\cite{Zannier2019} The InAs/InP NWs are grown by Au-assisted chemical beam epitaxy, using Au nano\-particles obtained by thermal de-wetting of a thin Au film on an InAs (111)B substrate. The NW is comprised of two $5.5$\,nm long segments of InP separated by $19$\,nm in an otherwise wurtzite InAs NW. The InP segments act as hard-wall tunnel barriers for electrons, since they induce a conduction band edge offset of $400$\,meV at the two atomically sharp InAs/InP interfaces, resulting in the formation of a QD between the two InP segments.

In this publication, we discuss two devices (\rm{I} and \rm{II}), both fabricated on a degenerately p-doped silicon substrate acting as a global back\-gate with a $400$\,nm thick SiO$_{\rm 2}$ capping layer. In device {\rm I}, the source contact to the NW is made of a titanium/gold film with a thickness of $5$/$65$\,nm whereas the drain contact is made of a titanium/aluminum film with a thickness of $5$/$55$\,nm. In device {\rm II}, both source and drain electrical contacts to the NW are made of titanium/gold films. Device~{\rm I} incorporates, in addition to a back\-gate, also a side\-gate, while this is absent in device~{\rm II}. In both devices, the native oxide of the NWs is removed with an (NH$_4$)$_2$S$_x$:H$_2$O solution before evaporation of the metal contacts.~\cite{Suyatin2007} In this work, we study transport at energies above the  superconducting gap of bulk Al ($\Delta \approx 210$\,$\mu$eV)~\cite{Court2007} and the discussion of the induced superconducting effects are beyond the scope of this paper.

\begin{figure}[]
\centering
\includegraphics[width=\columnwidth]{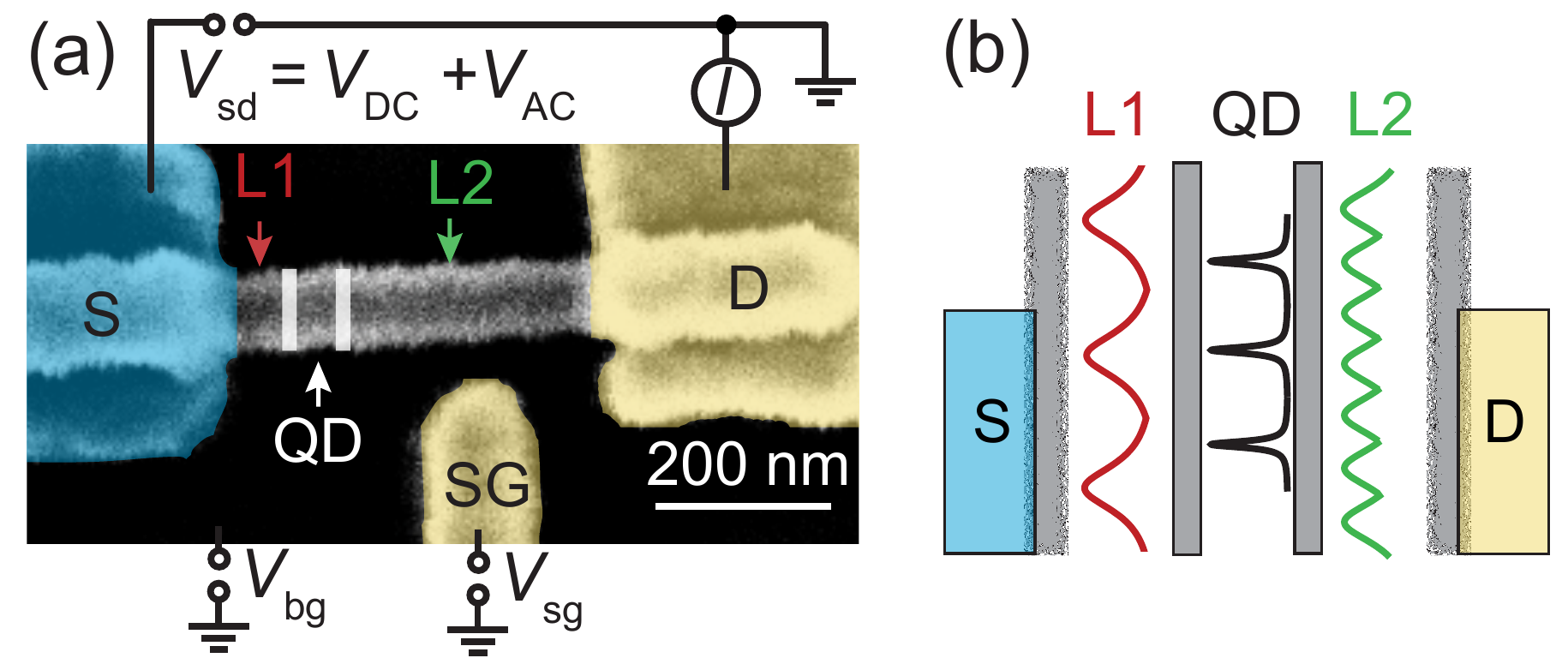}
\caption{(a) False color scanning electron microscopy (SEM) image of device \rm{I} that consists of an InAs NW with two short InP segments (white) that form an epitaxially defined quantum dot (QD). Between the epitaxially defined QD and the source S (blue) and drain D (yellow) contacts are bare InAs segments to which we refer to as the NW lead segments L1, and L2. Adjacent to the NW lead on L2 at the drain side, a local side\-gate is placed. In addition, the substrate acts as a global back\-gate. (b) Illustration of the sharp density of states of the QD and modulated density of states in the two NW lead segments L1 and L2. Note, that we indicated the barriers between source contact and L1, and between drain contact and L2, by a fuzzy box to indicated that in practice this barriers is ill defined. This is in contrasts to the barriers forming the QD. They are precisely known, both in position and size.}
\label{fig1}
\end{figure}


A false color scanning electron microscopy (SEM) image of device {\rm I} is shown in Fig.~\ref{fig1}(a). Here, the location of the epitaxial QD (white), as well as the two bare InAs NW lead segments L1, L2 are indicated. The chemical potential of the QD, L1 and L2 are tuned simultaneously by the back\-gate $V_{bg}$ and the side\-gate voltage $V_{sg}$, but with different lever arms. As we will show, the larger lever arms of the side\-gate allows to clearly discriminate a lead-state from a QD resonance.

The electrical measurements were performed in a dilution refrigerator at an electron temperature of $50$\,mK.~\cite{Thomas2020} The source-drain voltage bias $V_{sd}$ is applied to the source contact S, while  the current $I$ is measured at the grounded drain contact D. Note, in the following we use both upper- and lowercase symbols for source and drain when appropriate. The differential conductance $dI/dV = I_{\textrm{AC}}/V_{\textrm{AC}}$ is measured using standard lock-in techniques with an excitation voltage $V_{\textrm{AC}}=4$\,$\mu$V as a function of (the \textrm{DC}-part of) $V_{sd}$ and $V_{bg}$ or $V_{sg}$.

Figure~\ref{fig1}(b) shows a sketch of the local density of states (DOS) in the system including: the source (blue) and drain (yellow) metal contacts, the two lead segments L1 (red) and L2 (green), and the epitaxial QD (black). The modulated DOS in L1 and L2 originates from zero-dimensional (0D) confinement as a result of the epitaxially-defined InP hard-wall tunnel barriers on either side of the QD
%
and an effective barrier at the interface between the InAs and the metal contacts. The exact potential profile between the semiconductor and the metal is not known since it depends on various factors: the work functions of both materials, charge trapped at the interface, and the doping state in the NW. However, as is shown by the experiment, there is a potential step leading to a finite probability for electrons to be reflected back at the vicinity of the InAs-metal interface. Since we aim at low-ohmic contacts ($\ll h/2e^{2}$) and have applied the sulfur-etching technique, which is known to leave the InAs in a highly electron-doped state, the 0D-confinement is expected to be very asymmetric with the life-time mostly determined by the transparency of the metal contact and not by the tunnel coupling between the lead segment and the integrated QD. Such overlapping `quasi'-0D states lead to the DOS modulation. The spacing of the DOS peaks in the lead
segment DoS is then inversely proportional to the length of the leads (for a given Fermi velocity), while the width scales inversely to the life time, determined by the InAs contact transparency. Two possible configurations of the lead DOS in L1 and L2 are illustrated in Fig.~\ref{fig1}(b). In L1, the bound states weakly overlap, leading to a strong modulation of the lead DOS, while in L2 the bound states strongly overlap leading to a weak modulation of the lead DOS.

\section{Results and Discussions}
%
\begin{figure*}[]
\centering
\includegraphics[scale=1]{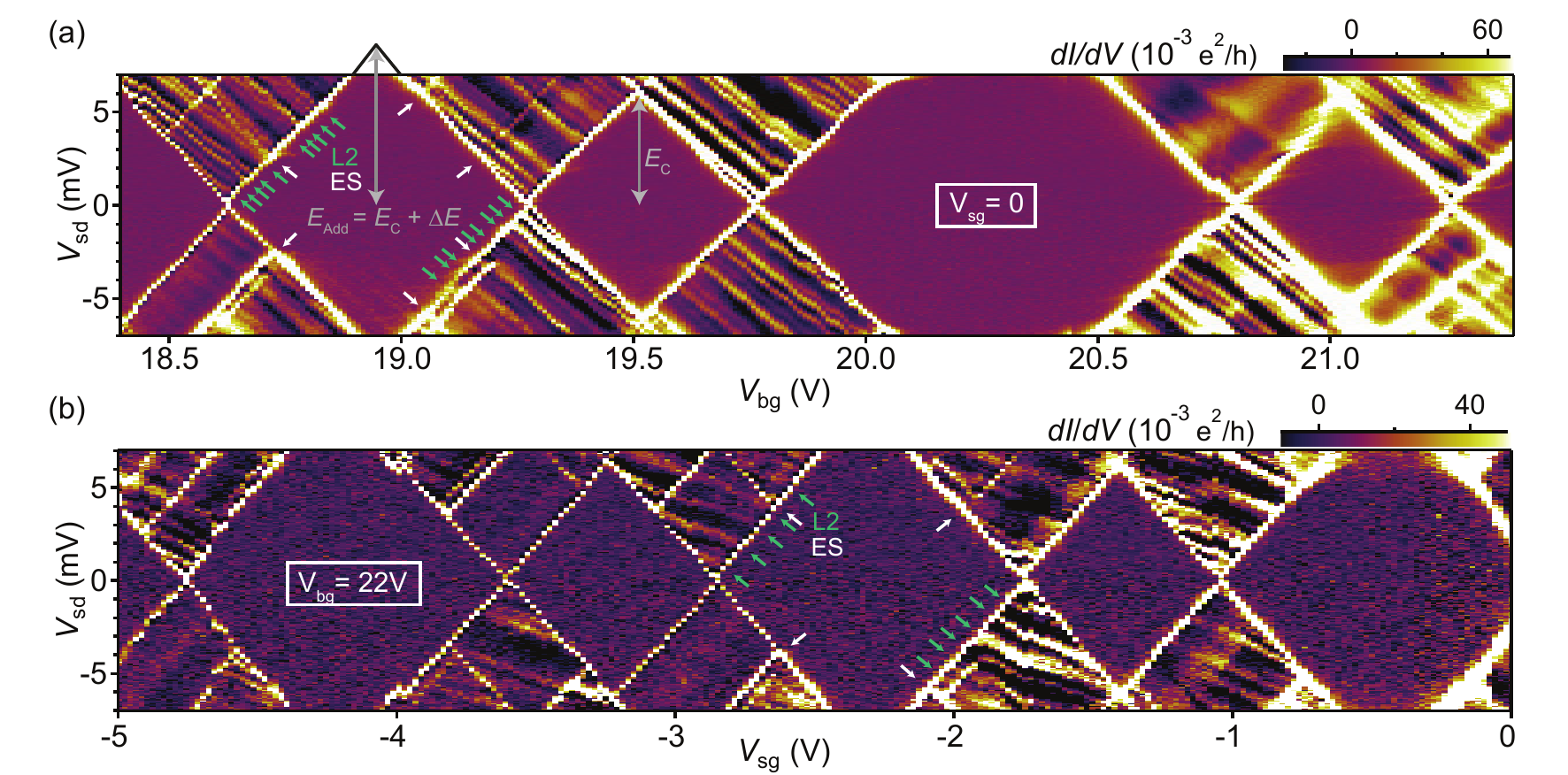}
\caption{(a) Differential conductance $dI/dV$ as a function of the voltage bias $V_{sd}$ (vertical axis) and back\-gate voltage $V_{bg}$ (horizontal axis) at $V_{sg}=0$\,V and (b) $dI/dV$ for side\-gate voltage $V_{sg}$ at $V_{bg} = 22$\,V. White arrows indicate excited state (ES) and green arrows lead state resonances originating from the L2 lead segment. Black arrows show how the addition energy for the first ($E_{\textrm{add}}$) and second (charging energy $E_C$) electron in an orbital is estimated. Form these values the orbital energy spacing $\Delta E$ can be estimated. }
\label{fig2}
\end{figure*}

Figure~\ref{fig2}(a) shows $dI/dV$ as a function of $V_{sd}$ and  $V_{bg}$.
The basic transport characteristics of a single QD are well understood.~\cite{Kouwenhoven1997}
In the two dimensional plane of $V_{sd}$ and $V_{bg}$, resonant tunneling through QD states gives rise to a diamond shape pattern. In the  Coulomb blockade region inside the diamonds, first order tunneling transport is  suppressed and the electron population on the QD is fixed. Outside of the diamonds, first order single electron tunneling is allowed. Here, additional conduction resonances appear  when a second transport channel for single electron transport via an excited state is available, indicated by the white arrows in one region in Fig.~\ref{fig2}(a).

In addition to excited state resonances, multiple resonances attributed to the  modulation of the DOS in the NW leads are highlighted by green arrows in Fig.~\ref{fig2}(a). Conduction features related to the lead-states are commonly seen in NW QD devices.~\cite{Bjork2003,Bjork2004,Bjork2005,Nilsson2016} However, to distinguish excited states from lead-state resonances is not always straightforward. In this device, the lead resonances with negative slopes are pronounced, whereas the lead resonances with positive slopes are hardly visible and if visible, much broader compared to the negative-sloped resonances. One obvious difference here is the different distances between the integrated QD and the source and drain metal contacts, which are $\sim\,50$\,nm and $\sim\,300$\,nm, respectively, and the different contact metals Al and Au. Both aspects can affect the lead resonances. The broad resonances could be an indication that the wurtzite InAs forms a lower barrier to Al than to Au.

In strongly confined QDs, the excited state (ES) resonances can sometimes be identified by relating the onset energy of the ES resonances to the energy spacing $\Delta E$ of the associated QD eigenstates estimated from the difference in addition energy for the larger $E_{\textrm{add}} = E_C + \Delta E$ and smaller $E_{\textrm{add}} = E_C$ diamonds, where $E_C$ is the charging energy, assumed to be constant.~\cite{Kouwenhoven1997} In Fig.~\ref{fig2}(a), the onset energy of the ES resonance (marked with ES) is  $\sim 2.45$~meV which agrees well with the estimated $\Delta E =2.5$\,meV with $E_C = 6.0$\,meV.

An alternative route to distinguish lead-states from ES resonances of the QD is by their slope $\Delta V_{sd}/\Delta V_{bg(sg)}$ in the $V_{sd}$ vs $V_{bg(sg)}$ plane, which is typically different to the slope of the resonances connected to the onset of resonant tunneling via QD states. In Fig.~\ref{fig2}(a), the difference in slope is not very strong. However, in Fig.~\ref{fig2}(b), where the differential conductance $dI/dV$ was recorded as a function of the side\-gate $V_{sg}$, the difference in the slopes $\Delta V_{sd}/\Delta V_{sg}$ is clearly visible. In the next section, we will derive general expressions for the various observable slopes.

\subsection*{The slopes of the resonance lines}

\begin{figure}
\centering
\includegraphics[width=\columnwidth]{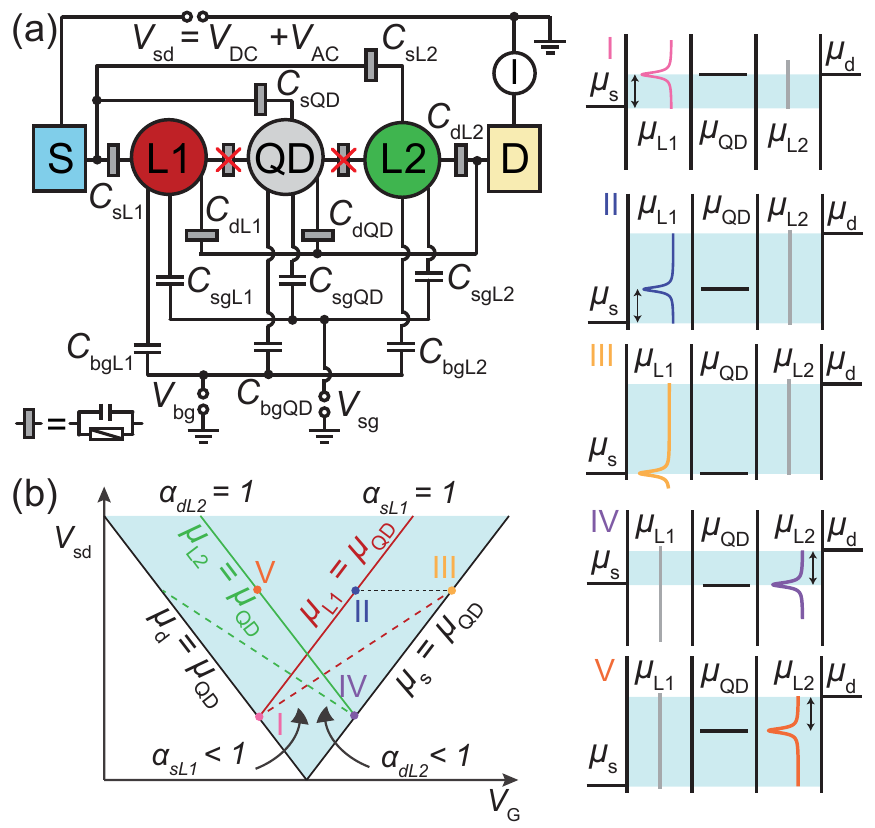}
\caption{(a) Equivalent circuit used to model the electrostatics of the device in Fig.~\ref{fig1}(a). Here, ``L1'' and ``L2'' are the left and right lead QDs and ``QD'' is the integrated QD. The gray boxes represent a tunnel coupling (resistor) and a capacitive coupling. The capacitors $C_{m,N}$ are labeled by two subscripts, where the first $m\in\{bg|sg|s|d\}$ refers to the metallic parts consisting of back\-gate and side\-gate, source, and drain and the second $N\in\{QD|L1|L2\}$ to the integrated QD and the lead segments L1 and L2.
(b) A stability diagram illustrating the resonances that arise when the electrochemical potential $\mu_{QD}$ in the integrated QD (sometimes loosely denoted as the `state' or `level' in the QD) aligns with either the electrochemical potential of the source $\mu_{s}$ or drain $\mu_{d}$ contacts (solid black). The lead-state resonances appear instead when $\mu_{QD}$ aligns with either $\mu_{L1}$ (red) or $\mu_{L2}$ (green). In the latter case, the solid lines represent the case when the lever arms $\alpha_{sL1}$ (red) or $\alpha_{dL2}$ (green) are equal to one, whereas for the dashed lines, $\alpha_{sL1}, \alpha_{dL2}< 1$. The alignment of the electrochemical potentials are depicted for five points indicated by the roman numerals. Here, the density of state for one lead shows a pronounced peak, while for the opposite lead an constant DOS is assumed.}
\label{fig3}
\end{figure}

To quantitatively describe the slope $\Delta V_{sd}/\Delta V_{bg(sg)}$ of the  lead resonances in the differential conductance, we consider a capacitor model consisting of three non-interacting QD elements, as depicted in Fig.~\ref{fig3}(a). Here, the integrated QD is  denoted ``QD'', whereas the QDs formed in the left and right lead segments are denoted L1 and L2, respectively. These notations are also used in the subscripts to denote the capacitances $C_{m,N}$, where the first index $m\in\{bg|sg|s|d\}$ denotes the metallic parts, i.e. either back\-gate, side\-gate, source, or drain. The second index $N\in\{QD|L1|L2\}$ denotes parts of the NW, i.e. the integrated QD and the lead segments L1 and L2. By non-interacting, we refer to the assumption that the capacitance between L1 and QD and between L2 and QD is neglected. This is justified by the geometrical smallness of both electrodes in these two cases. Furthermore, the charging energy on the QD and leads is not included in the model.


In the following we first derive a general expression for the slope of the relevant resonances. The conventional resonance of a single QD connected to source and drain with a constant DOS appears if $\mu_{QD}=\mu_{s,d}$. On the other hand, the main lead-state resonances that we observe arise when $\mu_{QD}=\mu_{L1,L2}$. Further resonances could be anticipated if $\mu_{L1,L2}=\mu_{s,d}$. However they are not observed in the experiment.

We start with the expressions for the electrochemical potentials of the lead segments $\mu_{L1}$, $\mu_{L2}$ and the integrated QD $\mu_{QD}$, which are given by
%
%
%
%
\begin{equation}\label{mu}
  \mu_N = q \sum_m \alpha_{mN} V_m \text{,}
\end{equation}
where $V_m\in\{V_{bg}|V_{sg}|V_s|V_d\}$ are the voltages applied to the back\-gate, side\-gate, source and drain electrodes and $q$ the charge of an electron, $q=-e$. $\alpha_{mN}$ are the lever arms, which are proportional to the respective coupling capacitances $C_{m,N}$. Here, we ignore possible offset charges for simplicity.
Since potentials are only defined up to an additive constant, all three equations in Eq.~\ref{mu} must remain valid if one adds the constant $\Delta E$ to $\mu_{L1,L2,QD}$ and to all four $qV_{bg,sg,s,d}$ values. Consequently, we obtain from Eq.~\ref{mu}:
%
%
%
\begin{equation}\label{sum_alpha}
  \sum_m \alpha_{mN}= 1 \text{.}
\end{equation}
A general expression then follows from Eq.~\ref{sum_alpha} for the lever arms in terms of the coupling capacitances:
\begin{equation}
  \alpha_{mN} = \frac{C_{mN}}{C_{\Sigma,N}},
\end{equation}
where $C_{\Sigma,N}=\sum_{m}C_{mN}$. The drain electrode is kept grounded throughout the experiments, thus $V_d = 0$. Furthermore, when a back\-gate sweep is shown, the side\-gate is at a constant potential and vice versa.

A sketch of the resonances due to onset of tunneling via the QD and lead-states is shown in Fig.~\ref{fig3}(b). A positive current will flow between source and drain if $\mu_s \leq  \mu_{QD}\leq \mu_d$ for positive bias and a negative one if $\mu_s \geq  \mu_{QD}\geq \mu_d$ for negative bias. The resonances constituting the edges of the diamond pattern in Fig.~\ref{fig2} appear when $\mu_{QD}$ is aligned with  $\mu_{d}$ or $\mu_{s}$, indicated in black in Fig.~\ref{fig3}(b). This condition then reads for positive and negative bias:
\begin{equation} \label{crit1}
  \mu_{s,d} = \mu_{QD}.
\end{equation}
Note, that $\mu_{s}=\mu_{0}+qV_{sd}$ and $\mu_{d}=\mu_{0}$, where $\mu_{0}$ is a constant value that does not enter the slopes. As a basis for deriving the slopes of the lead resonances, we postulate that the resonances associated with the lead-states appear in $dV/dI$, if a state in L1 or L2 is in resonance with a state on the QD and at the same time within the bias window. Thus, for positive $V_{sd}$ the condition reads:
\begin{equation} \label{crit2}
  \mu_{s} \leq \mu_{L1(2)} = \mu_{QD}\leq \mu_{d}
\end{equation}
and an analog condition for negative voltage bias. Taking all conditions formulated in Eqs.~\ref{crit1}-\ref{crit2}
yields four slopes for each two-dimensional plot of the $dI/dV$ as a function of $V_{sd}$ and $V_{bg}$ and additionally as function of $V_{sd}$ and $V_{sg}$. So there are eight measurable slopes in the experiment for device~{\rm 1}, which we could take as measures for the lever arms. In addition, there are three normalization conditions formulated in Eq.~\ref{sum_alpha}. We find that out of these $11$ equations, one is not independent and hence only $10$ unknowns can be obtained. However, in our model there are in total $12$ lever arms $\alpha_{mN}$, as seen from Eq.~\ref{mu} and as can be counted from the number of capacitors in Fig.~\ref{fig3}(a). Therefore, we have to make in addition two assumptions in our circuit model to obtain a well formulated problem. It is plausible to assume that the lever arms from source to the opposite lead 2 and from drain the other opposite lead 1 are small. Therefore, we set $\alpha_{sL2} = 0$ and $\alpha_{dL1} = 0$.

We define the slopes $S_{g(c)}$ of the resonance lines by two indices. The first one $g\in\{bg,sg\}$ refers to whether we look at the slopes in the $V_{sd}$ versus $V_{bg}$ or in the $V_{sd}$ versus $V_{sg}$ plane. The second index $(c)$ refers to the applied resonance condition. This can read, for example, as $S_{g(S=QD)}$, $S_{g(D=L2)}$ or $S_{g(L1=QD)}$, where $g\in\{bg,sg\}$. This can be understood as a QD state aligned with the electrochemical potential of the source, a state in lead L2 in resonance with the drain potential, or a state of the QD in resonance with a state in L1, respectively. In principle, there are in total $8$ possible slopes per gate voltage.

The slopes for the QD resonance with respect to source $S_{bg(S=QD)}$ and with respect to drain $S_{bg(D=QD)}$ for back\-gate sweeps can be expressed in terms of the lever arms using Eq.~\ref{mu} and Eq.~\ref{crit1} as (see Appendix):
\begin{align}
  S_{bg(S=QD)} &\equiv \left.\frac{\Delta V_{sd}}{\Delta V_{bg}}\right|_{\mu_s=\mu_{QD}} = \frac{\alpha_{bgQD}}{1-\alpha_{sQD}} \label{S_bg(S=QD)} \\
  S_{bg(D=QD)} &\equiv \left.\frac{\Delta V_{sd}}{\Delta V_{bg}}\right|_{\mu_d=\mu_{QD}} = -\frac{\alpha_{bgQD}}{\alpha_{sQD}} \label{S_bg(D=QD)},
 \end{align}
and analogous for side\-gate sweeps. These are formally four equations with three unknowns: $\alpha_{bg(sg)QD}$ and $\alpha_{sQD}$. Hence, one equation depends on the others.
In a similar way (see Appendix), the slopes of the lead resonances, for example for L1, $S_{bg(L1=QD)}$, are obtained by using the resonance criteria Eq.~\ref{crit2} together with Eq.~\ref{mu} for a backgate sweep:
\begin{equation} \label{S_bg(L1=QD)}
  S_{bg(L1=QD)} \equiv \left.\frac{\Delta V_{sd}}{\Delta V_{bg}}\right|_{\mu_{L1}=\mu_{QD}} = \frac{\alpha_{bgQD} - \alpha_{bgL1}}{\alpha_{sL1} - \alpha_{sQD}}
\end{equation}
and analogous for lead L2:
\begin{equation} \label{S_bg(L2=QD)}
  S_{bg(L2=QD)} \equiv \left.\frac{\Delta V_{sd}}{\Delta V_{bg}}\right|_{\mu_{L2}=\mu_{QD}} = \frac{\alpha_{bgQD} - \alpha_{bgL2}}{\alpha_{sL2} - \alpha_{sQD}},
\end{equation}
and analogous for side\-gate sweeps, for which the index $bg$ has to be replaced by $sg$.
Since we have $10$ equations and $10$ unknowns, one can now obtain all lever arms from the measured slopes taken from the experimental $dI/dV$ plots. The corresponding equations are given in the Appendix.

%

Now as an example, let us consider how the different lever arms affect the slope $S_{bg(L1=QD)}$. As seen in Eq.~\ref{S_bg(L1=QD)}, the slope is governed by the interplay between the lever arms to the epitaxial QD and the lead segment L1. We consider the case when the lever arm to the closest metal reservoir dominates, i.e. $\alpha_{sL1}\lesssim 1$. We now write $\alpha_{sL1} = 1-\epsilon$, where $0< \epsilon \ll 1$ denotes all other leaver arms to L1, $(\alpha_{bgL1}+\alpha_{sgL1})\ll 1$. It follows that $\alpha_{bgL1}\leq \epsilon$ and $\alpha_{sgL1}\leq \epsilon$.
From Eq.~\ref{S_bg(L1=QD)} we now obtain $S_{bgL1}=(\alpha_{bgQD}-\alpha_{bgL1})/(1-\epsilon-\alpha_{sQD})$. This expression converges to
$\alpha_{bgQD}/(1-\alpha_{sQD})$ for $\epsilon\rightarrow 0$, which is identical to Eq.~\ref{S_bg(S=QD)}. Hence, $S_{bg(L1=QD)}\rightarrow S_{bg(S=QD)}$ for $\alpha_{sL1}\rightarrow 1$. The slope of the L1 resonance approaches the positive slope of the QD resonance relative to the source potential if the coupling of L1 to S dominates.
%
%
%
This can be understood by recalling that the $S_{bg(S=QD)}$-resonance originates from the condition $\mu_{QD} = \mu_{S}$. If $\alpha_{sL1}$ dominates, $\mu_{L1}$ will follow $\mu_{S}$ as indicated with the solid green line in Fig.~\ref{fig3}(b). That $\mu_{L1}$ follows $\mu_{S}$ is also indicated with the arrow in the schematic of the alignment of the electrochemical potentials for the point indicated with the Roman numerals I and II.
Considering L2, one finds analogously that the slope $S_{bg(L2=QD)}$ approaches the negative slope $S_{bg(D=QD)}$ of the QD when $\alpha_{dL2}$ is dominating. Then, $\mu_{L2}$ is following $\mu_{D}$, as is the case for the solid red line in Fig.~\ref{fig3}(b).

In the experiments where the back\-gate is swept (Fig.~\ref{fig2}(a)), we often see lead resonances with similar slopes as the QD resonances, i.e. $S_{bg(S=QD)}$ and $S_{bg(D=QD)}$ defining the bias window. This indicates that the capacitance between the leads and the back\-gate is considerably smaller than the capacitance between L1 and source and L2 and drain. In contrast, if we consider side\-gate sweeps, the slope corresponding to the L2 resonance, $S_{sg(L2=QD)}$, can deviate considerably from the slope $S_{sg(D=QD)}$ of the QD resonance relative to $\mu_D$ due to a relatively large lever arm $\alpha_{sgL2}$. This situation is depicted with the red and green dashed lines in Fig.~\ref{fig3}(b) and seen experimentally in Fig.~\ref{fig2}(b).

As introduced before, it could be possible to observe up to $16$ families of slopes. To discuss why the majority of them are not observed in the experiment, let us briefly consider, as an example, the slope $S_{bg(L1=S)}=\alpha_{bgL1}/(1-\alpha_{\textrm{SL1}})$. Since we observe in general that the lever arm between source and L1 is large, we write $\alpha_{SL1}=1-\epsilon^*$, resulting in $\alpha_{bgL1}+\alpha_{sgL1}=\epsilon^*$. The experimental data in Fig.~\ref{fig2} yields approximately $\alpha_{sgL1}\approx 2 \alpha_{bgL1}$ (see Table \ref{tab1}) and therefore $\alpha_{bgL1}\simeq\epsilon^*/3$. Taking this together then yields for the slope $S_{bg(L1=S)}\simeq 1/3$. If we compare with the experiment where the slopes have typical values of $0.01$ (see Table \ref{tab1} and \ref{tab2}), $S_{bg(L1=S)}$ is larger by an order of magnitude. Since this resonance line is mostly vertical, it will appear much weaker in a differential conductance measurement and this may explain why this slope is not seen. One can also estimate in a similar fashion the opposite slope $S_{bg(L1=D)}$ for which one obtains $\simeq\epsilon^*/3$. For $\epsilon^*\rightarrow 0$, this slope disappears. In practice, the slope is small, but not much smaller than $S_{bg(QD=D)}$ and one could therefore expect to observe such resonant lines as well. These lines should not yield a negative $dI/dV$ (NDC). In contrast, all observed lead-state resonance lines are accompanied by a negative `wing' of $dI/dV$ when plotted in the $V_{sd}$ versus $V_{g}$ plane. This suggests that these are all due to the resonance condition $\mu_{L1 (L2)}=\mu_{QD}$. The NDC appears if, the lead-state chemical potential $\mu_{L1}$ ($\mu_{L2}$) shifts at a different rate as the QD resonance $\mu_(QD)$ as a function of $V_{sd}$. This then results in the condition $\alpha_{sQD}\not = \alpha_{sL1}$ for lead state $L_1$ and $\alpha_{sQD}\not = \alpha_{sL2}$ for lead state $L_2$. This condition holds in practice, since otherwise the slopes in Eq.~\ref{S_bg(L1=QD)} and Eq.~\ref{S_bg(L2=QD)} would be singular.
\subsection*{Resonant tunneling model}

In order to substantiate the postulate that resonances appear in $dI/dV$ when $\mu_{L1(L2)} = \mu_{QD}$, we use a resonant tunneling model to calculate the current $I$ through the three independent weakly tunnel coupled QDs depicted in Fig.~\ref{fig3}(a)~\cite{Beenakker1991}
\begin{equation} \label{I}
I(V_{sd},V_{g}) = \frac{e}{h}\int T_{total}(E)[f_s(E)-f_d(E)]dE
\end{equation}
where $f_{s/d}(E) = 1/(1 + \exp((E-\mu_{s/d})/k_B T)$ are the Fermi-Dirac distributions in the source and drain contacts. The total transmission function is taken as the product of the individual transmission functions of the three segments $T_{N}(E)$, $N\in\{QD|L1|L2\}$):
\begin{equation} \label{T1T2T3}
  T_{total}(E) = T_{L1}(E)T_{QD}(E)T_{L2}(E) \text{.}
\end{equation}
In the case of the epitaxial QD, the transmission function is taken as the Breit-Wigner (BW) function: $T_{QD}(E) = A_{QD}\Gamma_{QD}^2 /[(E-\mu_{QD})^2 + \Gamma_{QD}^2]$, where   $\Gamma_{QD}$ is the sum of the tunnel couplings of the two InP  barriers defining the QD and the amplitude $A_{QD}$ accounts for the asymmetry between the tunnel couplings. $E-\mu_{QD}$ is the detuning from the electrochemical potential of the QD.

In the case of the lead-states, the BW transmission functions $T_{L1(2),i}(E) = A_{L1(2),i}\Gamma_{L1(2),i}^2 /[(E-\mu_{L1(2),i}-E_{\textrm{orb},i})^2 + \Gamma_{L1(2),i}^2]$ is shifted by an additional term $E_{\textrm{orb},i}$ that models the energy spacings of the different states $i$ contributing to the lead and is extracted from the point where the lead  resonance meets the diamond edge.  The contribution to the transmission probability of the lead resonances identified in the experimental data are summed up and an offset value $T_{\textrm{off}}$ is added to capture off-resonance transport: $T_{L1(2)}(E) = T_{\textrm{off}} +\Sigma_i T_{L1(2),i}(E)$. Off-resonance transport includes for instance higher order tunnelling process, e.g. cotunneling and  possible broad ``background'' states that are poorly gate tuned. In addition, asymmetric line shapes might occur due to Fano interference~\cite{Bar-Ad1997} not captured in the BW transmission function.

\begin{figure}[]
\includegraphics[scale=1]{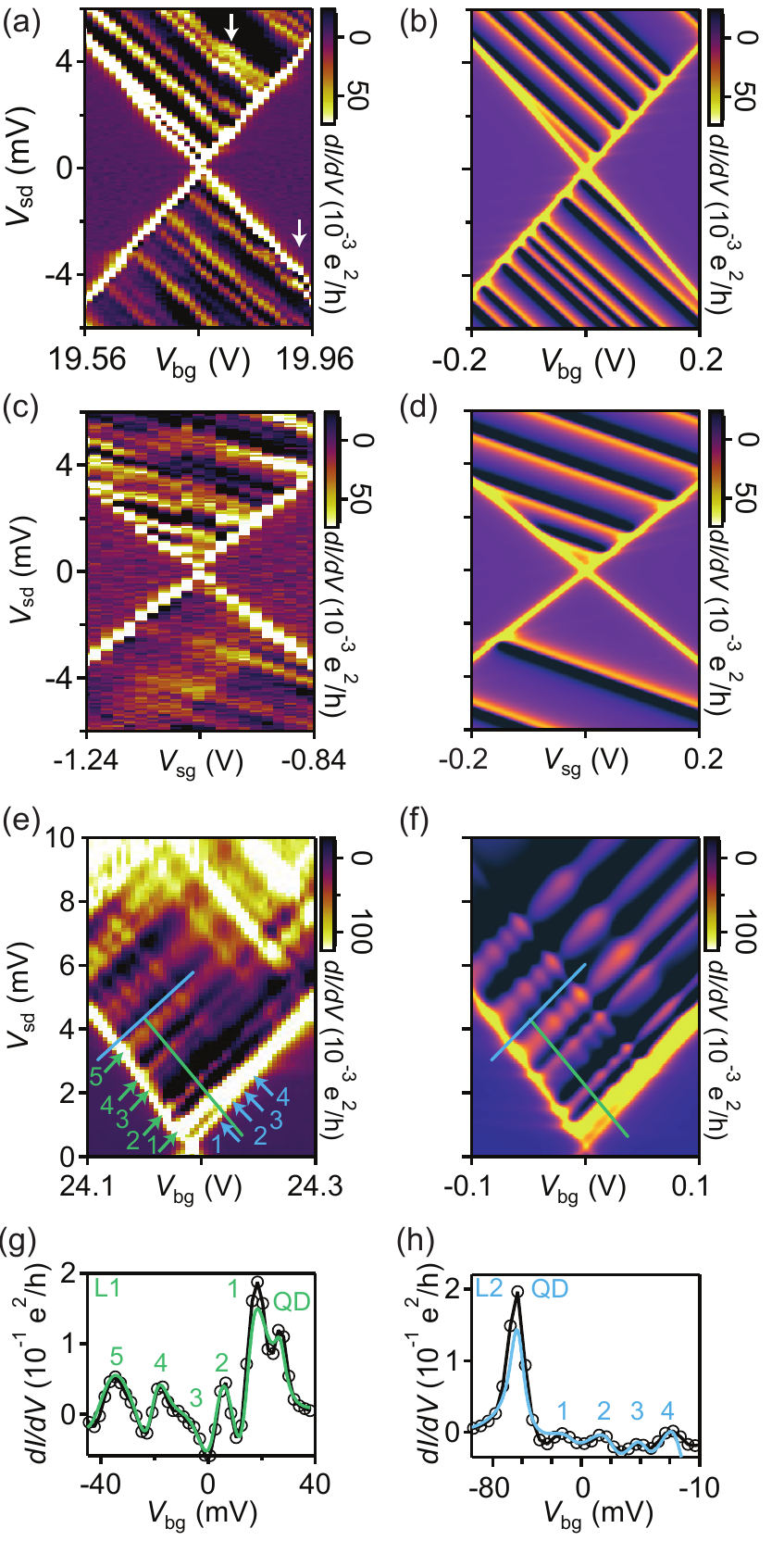}
\caption{Differential conductance $dI/dV$ as a function of voltage bias $V_{sd}$ and (a) back\-gate voltage $V_{bg}$ and (c) side\-gate voltage $V_{sg}$ for device \textrm{I}. Panel (b) and (d) show the modelled $dI/dV$ where the lever arms were extracted from the slopes of the measured data seen the panel (a) and (c). (e) Shows    $dI/dV$ as a function of $V_{sd}$ and  $V_{bg}$ measured for device \textrm{II}. The lead resonances are indicated with green (L1) and blue (L2) lines and arrows. (f) Results from the model with input values extracted from (e). (g) and (f) show line-cuts along the green and blue curves in (e) and (f), respectively. Here, the solid coloured lines are the model values and the black markers are the experimental data.}
\label{fig4}
\end{figure}

We extract all the lever arms form the experimental data, i.e. from the measured slopes according to the procedure outlined before. Then we plot $dI/dV$ using the resonant tunneling model (Eq.~\ref{I} and Eq.~\ref{T1T2T3}) with the tunnel couplings and asymmetry factors as fitting parameter. Even without fine-tuning of the coupling and asymmetry parameters, this already provides a good picture of the $dI/dV$ as seen in Fig.~\ref{fig4}.

\begin{table}[!h]
  \caption{ Lever arms $\alpha_{mN}$ for the epitaxial QD and the lead  segment (L2) extracted from the slopes $S_{g(c)}$ of the experimental data in Fig.~\ref{fig4}(a) and Fig.~\ref{fig4}(c) of device \textrm{II}.} \label{tab1}
    \begin{ruledtabular} 
      \begin{tabular}{m{0.1\textwidth} p{0.175\textwidth} p{0.175\textwidth} }
        \rule{0pt}{3ex} & QD  & L2  \\[2pt]
        \hline
        \rule{0pt}{3ex} pos. slope  & $S_{bg(S=QD)}=0.023$ \newline $S_{sg(S=QD)}=0.017$ & -  \\
        \rule{0pt}{3ex} neg. slope  & $S_{bg(D=QD)}=-0.025$ \newline $S_{sg(D=QD)}=-0.017$ & $S_{bg(L2=QD)}= -0.020 $  \newline$S_{sg(L2=QD)}=-0.007$  \\
        \rule{0pt}{3ex} lever arms  & $\alpha_{bgQD}= 0.012$ \newline $\alpha_{sgQD}= 0.008$  \newline $\alpha_{sQD}= 0.491$ \newline $\alpha_{dQD}= 0.489$ & $\alpha_{bgL2}= 0.002$ \newline $\alpha_{sgL2}=0.005$ \newline $\alpha_{sL2}= 0$ \newline $\alpha_{dL2}=0.993$
\end{tabular}
\end{ruledtabular}
\end{table}

Figure~\ref{fig4}(a) and Fig.~\ref{fig4}(c) show  regions of Fig.~\ref{fig2}(a) and Fig.~\ref{fig2}(b) along with the corresponding results from the modelling in Fig.~\ref{fig4}(b) and Fig.~\ref{fig4}(d). Only lead resonances with negative slope have been included in the model, since the lead resonances with positive slope are difficult to discern in the experimental data. The lever arms to the lead segments used in the model are extracted from the average of the slopes in the experiment (see Table~\ref{tab1}). Figure~\ref{fig4}(b) and Fig.~\ref{fig4}(d) capture the slopes of the resonances, where the asymmetry and Gamma were chosen to fit the data. We find that the model $dI/dV$ shows strong negative $dI/dV$ in very good agreement with the experiment. One interesting difference between the experiment and the model are sharp shifts in the resonance positions indicated by arrows. We speculate that these jumps are due to charging effects in the shorter lead L1 affecting the electrostatics in the system, which were not taken into consideration in the model. If single electron tunneling is considered, the electro\-chemical potentials $\mu_{N}$ in Eq.~\ref{mu} are not linear in the applied voltages, but also depend on the number of charges accumulated in the two lead segments and the QD.

\begin{table*}[]
 \caption{Lever arms for the epitaxial QD and for the lead  segments (L1,L2) used to model Fig.~\ref{fig4}(f), calculated from the extracted values of the slopes from the experimental data from Fig.~\ref{fig4}(e). Tunnel couplings and asymmetry factors resulting from the fit of the cross-sections in Fig.~\ref{fig4}(g) and Fig.~\ref{fig4}(h) of device \textrm{II}. } \label{tab2}
\begin{ruledtabular} 
  \begin{tabular}{m{0.25\textwidth} p{0.25\textwidth} p{0.25\textwidth} p{0.25\textwidth} }
    \rule{0pt}{3ex}   & QD  & L1 & L2  \\ [2pt]
    \hline
    \rule{0pt}{3ex} Positive slope & $S_{bg(S=QD)}=0.03940$ & $S_{bg(L1=QD)}=0.03555$ & - \\
    \rule{0pt}{3ex} Negative slope & $S_{bg(D=QD)}=-0.04810$  & -  & $S_{bg(L2=QD)}=-0.04157$   \\
    \rule{0pt}{3ex} Lever arms & $\alpha_{bgQD}=0.0217$ \newline $\alpha_{sQD}= 0.4503$ \newline $\alpha_{dQD}= 0.5281$ & $\alpha_{bgL1}=   0.0022 $ \newline $\alpha_{sL1}=0.9978$ \newline $\alpha_{dL1}=0$ & $\alpha_{bgL2}=0.0029$ \newline $\alpha_{sL2}=0$ \newline $\alpha_{dL2}=0.9971$ \\
    \rule{0pt}{3ex} Tunnel couplings \newline ($\mu$eV)  & $\Gamma_{QD}=110$ & $\Gamma_{L1,1}=175$ \newline $\Gamma_{L1,2}=200$ \newline $\Gamma_{L1,3}=475$ \newline $\Gamma_{L1,4}=150$ \newline $\Gamma_{L1,5}=450$ & $\Gamma_{L2,1}=125$ \newline $\Gamma_{L2,2}=150$ \newline $\Gamma_{L2,3}=110$ \newline $\Gamma_{L2,4}=110$ \\
    \rule{0pt}{3ex} Asymmetry factors  & $A_{QD}=0.7$ & $A_{L1,1}=0.8$ \newline $A_{L1,2}=0.58$ \newline $A_{L1,3}=0.29$ \newline $A_{L1,4}=0.34$ \newline $A_{L1,5}=0.62$ & $A_{L2,1}=0.07$ \newline $A_{L2,2}=0.17$ \newline $A_{L2,3}=0.14$ \newline $A_{L2,4}=0.32$
  \end{tabular}
\end{ruledtabular}
\end{table*}

Next, we go one step further and use the model to estimate the tunnel couplings and the asymmetry factors of the lead-states by using them as fit parameters to reproduce cross-sections in the experimental data. In Fig.~\ref{fig4}(e), experimental data from device~\textrm{II} are shown. This device is based on the same type of NW, however, the lead segments are more similar in length, $\sim 350$ and $\sim 600$\,nm for L1 and L2. Furthermore, the source and drain contacts are both made of the same material Ti/Au and there is no side\-gate. Device~\textrm{II} clearly displays lead resonances of both positive and negative slopes, highlighted by green and blue arrows. In addition, the width of the resonance for both L1 and L2 are similar, all reflecting the more symmetric nature of the device configuration.

Values for the extracted slopes and calculated lever arms together with the fitted tunnel couplings are summarized in Table~\ref{tab2}. As a starting point for the tunnel couplings and asymmetry factor for the epitaxial QD, values estimated from an in-depth study of this particular type of QDs were used.~\cite{Thomas2020}
The tunnel couplings and asymmetry factors of the leads together with $T_{\textrm{off}}$, were used as free fit parameters. As can be seen in Fig.~\ref{fig4}(g) and Fig.~\ref{fig4}(h) remarkably accurate fits can be obtained. The corresponding values in Table~\ref{tab2} show that the tunnel coupling of the epitaxial QD is slightly smaller compared to the lead segments: $\Gamma_{QD}\sim 100$\,$\mu$eV, whereas $\Gamma_L\simeq 100-500$\,$\mu$eV. This shows that the coupling of the lead segments to evaporated metal contacts is not very strong and on the order of the level spacing in the leads. In contrast to the data from device {\rm II}, device {\rm I} displays very asymmetric $\Gamma_L$ for L1 and L2. In sample {\rm I}, the coupling of the lead-state L1 to the source contact appears to be much wider, suggesting that aluminium provides a better contact to the conduction band in InAs compared to Au.

\section{Conclusions}

In conclusion, we have studied and identified lead-state resonances in semiconducting NWs with an integrated QD as spectrometer and we could show that they are due to the resonance condition $\mu_{L1 (L2)}=\mu_{QD}$. All lever arms can be obtained from the measured dependencies of the QD and lead-state resonances as a function of $V_{sd}$ and $V_{bg}$/$V_{sg}$ using a simplified electrostatic model if the capacitive coupling between source/drain to the opposite lead L2/L1 is negligible. It is also possible to simulate the full $dI/dV$ in a resonant tunneling model, expressing the resonances by effective tunnel couplings and asymmetry parameters. Overall, a very good agreement between the model and the experimental data can be achieved.

In measurements where the back\-gate is swept, the lead-state resonances have a slope that often deviates only slightly from the lines due to an onset of transport through the QD relative to the source and drain electrochemical potentials. This leads to difficulties distinguishing lead-state from QD resonances, which, however, can be easily discriminated using a more local sidegate.

We have also demonstrated, that in a typical gating configuration the lever arms between lead L1 (L2) and source (drain) contact dominate. Consequently, the electrochemical potentials of the leads follow closely the ones of the contacts, i.e. $\mu_{L1}\simeq\mu_s$ and $\mu_{L2}\simeq\mu_d$. Nonetheless, there is a difference between the lead-state slopes that are measured versus $V_{bg}$ and versus $V_{sg}$. This can be seen from Eq.~\ref{S_bg(L1=QD)}. The slope of, for example, lead-state L1 is proportional to the difference between the lever arm from the gate to the QD and the gate to L1: $S_{g(L1=QD)}\propto \alpha_{gQD}-\alpha_{gL1}$. Here, $g$ stands for either $bg$ or $sg$.
The slope is generally smaller when sweeping the sidegate since $\alpha_{gQD}$ is reduced and $\alpha_{gL1}$ is increased in that case.

We expect that for quantitative studies in which QDs will be used as spectroscopic tools and energy filters, accounting for the electrostatics of nanostructures will become increasingly relevant, since the device designs are becoming more complex and move towards real architectures, not the least in the field of quantum technologies.
\section{Appendix}
From the experiments, one obtains the different slopes of the resonance lines in the $V_{sd}$ versus $V_g$ plane(s) as ``input'' parameters. From these, one has to derive the lever arms $\alpha_{mN}$, which are the ``output'' parameters. We will do this explicitly in the following, assuming as before that there are two gates relevant in the device, the back\-gate and the side\-gate. The derivation with one gate alone is simpler, but follows in an analogous way.

We recall the definition of the lever arms $\alpha_{mN}$ given in Eq.~\ref{mu}. Here, the first index $m$ refers to an electrode, $m\in\{bg|sg|s|d\}$ and the second index $N$ to the three weakly-coupled systems, the QD and the left and right leads L1,L2: $N\in\{QD|L1|L2\}$. Since in our experiment $V_d=0$ we obtain explicitly the three equations for the chemical potentials $\mu_{N}$:
\begin{subequations}
  \begin{align}
    \mu_{QD}&=q\left(\alpha_{bgQD}V_{bg}+\alpha_{sgQD}V_{sg}+\alpha_{sQD}V_{s}\right)  \label{muQD}  \\
    \mu_{L1}&=q\left(\alpha_{bgL1}V_{bg}+\alpha_{sgL1}V_{sg}+\alpha_{sL1}V_{s}\right)  \label{muL1}  \\
    \mu_{L2}&=q\left(\alpha_{bgL2}V_{bg}+\alpha_{sgL2}V_{sg}+\alpha_{sL2}V_{s}\right)  \label{muL2}
  \end{align}
\end{subequations}
We now also explicitly express the three normalization equations from Eq.~\ref{sum_alpha}, recalling that $\alpha_{dL1}=0$ and $\alpha_{sL2}=0$. These conditions were necessary to have uniquely defined lever arms:
\begin{subequations}
  \begin{align}
    1&=\alpha_{bgQD}+\alpha_{sgQD}+\alpha_{sQD}+\alpha_{dQD} \label{sum_alphas_QD}  \\
    1&=\alpha_{bgL1}+\alpha_{sgL1}+\alpha_{sL1} \label{sum_alphas_L1}  \\
    1&=\alpha_{bgL2}+\alpha_{sgL2}+\alpha_{dL2} \label{sum_alphas_L2}
  \end{align}
\end{subequations}
Let us look first at the QD resonances in the $V_{sd}$ versus $V_{bg}$ plane. In such an experiment, $V_{sg}$ is constant. The slope of the resonance is the derivative of $V_{sd}$ with respect to $V_{bg}$ under the condition that $\mu_{QD}$ is aligned with either $V_{s}$ or $V_{d}=0$. Since $\mu_{s}=\mu_{0}+qV_s$, we obtain from Eq.~\ref{muQD} for $\mu_s=\mu_{QD}$:
\begin{equation}
  \Delta V_s=\alpha_{bgQD}\Delta V_{bg} + \alpha_{sQD}\Delta V_s  .
\end{equation}
Solving for the slope (with $V_{s}=V_{sd}$), yields:
\begin{equation}
  S_{bg(S=QD)} = \left.\frac{\Delta V_{sd}}{\Delta V_{bg}}\right|_{\mu_s=\mu_{QD}} = \frac{\alpha_{bgQD}}{1-\alpha_{sQD}}. \label{S_bg(S=QD)b}
\end{equation}
This is Eq.~\ref{S_bg(S=QD)}. If $\mu_{QD}$ is instead aligned with $\mu_d$, we obtain again from from Eq.~\ref{muQD}
\begin{equation}
  0=\alpha_{bgQD}\Delta V_{bg} + \alpha_{sQD}\Delta V_s ,
\end{equation}
and solving for the slope, yields:
\begin{equation}
  S_{bg(D=QD)} = \left.\frac{\Delta V_{sd}}{\Delta V_{bg}}\right|_{\mu_d=\mu_{QD}} = -\frac{\alpha_{bgQD}}{\alpha_{sQD}}. \label{S_bg(D=QD)b}
\end{equation}
We see that we have two slopes  $S_{bg(S=QD)}$ and $S_{bg(D=QD)}$ and two unknowns $\alpha_{bgQD}$ and $\alpha_{sQD}$. Hence, we can solve for these two $\alpha$'s. The same procedure can be done for the slopes measured in the $V_{sd}$ versus $V_{sg}$ plane. Here, the equations are:
\begin{subequations}
  \begin{align}
    S_{sg(S=QD)} &= \left.\frac{\Delta V_{sd}}{\Delta V_{sg}}\right|_{\mu_s=\mu_{QD}} = \frac{\alpha_{sgQD}}{1-\alpha_{sQD}} \text{ and} \label{S_sg(S=QD)b} \\
    S_{sg(D=QD)} &= \left.\frac{\Delta V_{sd}}{\Delta V_{sg}}\right|_{\mu_d=\mu_{QD}} = -\frac{\alpha_{sgQD}}{\alpha_{sQD}}. \label{S_sg(D=QD)b}
  \end{align}
\end{subequations}
We additionally obtain the pair $\alpha_{sgQD}$ and $\alpha_{sQD}$. We see that we obtain $\alpha_{sQD}$ twice. In theory, they should come out the same. Due to measurements errors, they can slightly deviate from each other in practice, in which case the average value is taken.
Now, the set $\{\alpha_{bgQD}, \alpha_{sgQD}, \alpha_{sQD}\}$ is already determined. The final lever arm to the QD, $\alpha_{dQD}$ is then obtained from the normalization condition Eq.~\ref{sum_alphas_QD}.

Next, we look at the lead-state resonances. We treat the two leads separately and start with the resonance $\mu_{L2}=\mu_{QD}$. From Eq.~\ref{muQD} and Eq.~\ref{muL2} we obtain in the $V_{sd}$ versus $V_{bg}$ plane:
\begin{equation}
  \alpha_{bgQD}\Delta V_{bg}+\alpha_{sQD}\Delta V_s=\alpha_{bgL2}\Delta V_{bg}+\alpha_{sL2}\Delta V_s.
\end{equation}
This yields for the slope $S_{bg(L2=QD)}$:
\begin{equation}
  S_{bg(L2=QD)}=\left.\frac{\Delta V_{sd}}{\Delta V_{bg}}\right|_{\mu_{L2}=\mu_{QD}}=\frac{\alpha_{bgQD}-\alpha_{bgL2}}{\alpha_{sL2}-\alpha_{sQD}}.  \label{S_bg(L2=QD)b}
\end{equation}
Similarly, in the $V_{sd}$ versus $V_{sg}$ plane one obtains:
\begin{equation}
  S_{sg(L2=QD)}=\left.\frac{\Delta V_{sd}}{\Delta V_{sg}}\right|_{\mu_{L2}=\mu_{QD}}=\frac{\alpha_{sgQD}-\alpha_{sgL2}}{\alpha_{sL2}-\alpha_{sQD}}.  \label{S_sg(L2=QD)b}
\end{equation}
Note, that in the last two equations, $\alpha_{sL2}=0$. For this reason, $\alpha_{bgL2}$ is uniquely determined by Eq.~\ref{S_bg(L2=QD)b} and, equally, $\alpha_{sgL2}$ by Eq.~\ref{S_sg(L2=QD)b}. The remaining lever-arm $\alpha_{dL2}$ to lead L2 is obtained from the normalization equation, Eq.~\ref{muL2}.

At this point, all lever arms to the QD and to lead L2 are known. Now we deduce the remaining lever arms to lead L1. Following Eq.~\ref{S_bg(L2=QD)b}, Eq.~\ref{S_sg(L2=QD)b} and the normalization condition Eq.~\ref{sum_alphas_L1} we obtain:
\begin{subequations}
  \begin{align}
    S_{bg(L1=QD)}&=\left.\frac{\Delta V_{sd}}{\Delta V_{bg}}\right|_{\mu_{L1}=\mu_{QD}}=\frac{\alpha_{bgQD}-\alpha_{bgL1}}{\alpha_{sL1}-\alpha_{sQD}},  \label{S_bg(L1=QD)b} \\
    S_{sg(L1=QD)}&=\left.\frac{\Delta V_{sd}}{\Delta V_{sg}}\right|_{\mu_{L1}=\mu_{QD}}=\frac{\alpha_{sgQD}-\alpha_{sgL1}}{\alpha_{sL1}-\alpha_{sQD}},  \label{S_sg(L1=QD)b} \\
    1&=\alpha_{bgL1}+\alpha_{sgL1}+\alpha_{sL1}.
  \end{align}
\end{subequations}
These are the remaining three equations for the remaining three unknowns $\alpha_{bgL1}$, $\alpha_{sgL1}$ and $\alpha_{sL1}$ (remember that $\alpha_{dL1}=0$). Solving for the three unknowns yields:
\begin{subequations}
  \begin{align}
    \alpha_{bgL1}&=\frac{S_{bg}(\alpha_{bgQD}+\alpha_{dQD})+\alpha_{bgQD}(S_{sg}-1)}{S_{bg}+S_{sg}-1} \\
    \alpha_{sgL1}&=\frac{S_{sg}(\alpha_{sgQD}+\alpha_{dQD})+\alpha_{sgQD}(S_{bg}-1)}{S_{bg}+S_{sg}-1} \\
    \alpha_{sL1} &=\frac{\alpha_{bgQD}+\alpha_{sgQD}+\alpha_{sQD}(S_{bg}+S_{sg})-1}{S_{bg}+S_{sg}-1}.
  \end{align}
\end{subequations}
Here, we have abbreviated the slopes with the symbols $S_{bg}$, $S_{sg}$. They both refer to slopes due to lead-state L1. All lever arms are thus determined. In the experiment it is possible that, for example, only resonances from lead-state L1 are visible. One can then still obtain all lever arms to L1, but there is no information on the lever arms to lead-state L2.

\subsection*{Author contributions}
F.T. and C.C. fabricated the device and F.T, C.C. and M.N. performed the measurements. M.N., F.T., and A.B. analyzed the data. C.J. supported the fabrication and discussions. M.N., F.T., A.B. and C.S. derived the equations. A.B. and F.T. worked out the resonant tunneling model F.R., V.Z., and L.S. developed the NW structure. C.S. and A.B. planned and designed the experiments, and participated in all discussions. All authors contributed to the manuscript.
\begin{acknowledgments}

This research was supported by the Swiss National Science Foundation through (a) a project grant entitled ‘Quantum Transport in Nanowires’ granted to CS (b) the National Centre of Competence in Research Quantum Science and Technology and (c) the Quant Era project SuperTop. It has further been supported by a PhD Grant from the Swiss Nanoscience Institute (SNI) and the University of Basel. This project has also received support from the European Union’s Horizon 2020 research and innovation programme under Grant agreement No. 828948, project AndQC and under the Marie Skłodowska-Curie actions (MSCA) COFUND by the project QUSTEC. The authors declare no competing financial interest. All data in this publication are available in numerical form, see: https://doi.org/10.5281/zenodo.4555268
\end{acknowledgments}


\begin{thebibliography}{34}%
\makeatletter
\providecommand \@ifxundefined [1]{%
 \@ifx{#1\undefined}
}%
\providecommand \@ifnum [1]{%
 \ifnum #1\expandafter \@firstoftwo
 \else \expandafter \@secondoftwo
 \fi
}%
\providecommand \@ifx [1]{%
 \ifx #1\expandafter \@firstoftwo
 \else \expandafter \@secondoftwo
 \fi
}%
\providecommand \natexlab [1]{#1}%
\providecommand \enquote  [1]{``#1''}%
\providecommand \bibnamefont  [1]{#1}%
\providecommand \bibfnamefont [1]{#1}%
\providecommand \citenamefont [1]{#1}%
\providecommand \href@noop [0]{\@secondoftwo}%
\providecommand \href [0]{\begingroup \@sanitize@url \@href}%
\providecommand \@href[1]{\@@startlink{#1}\@@href}%
\providecommand \@@href[1]{\endgroup#1\@@endlink}%
\providecommand \@sanitize@url [0]{\catcode `\\12\catcode `\$12\catcode
  `\&12\catcode `\#12\catcode `\^12\catcode `\_12\catcode `\%12\relax}%
\providecommand \@@startlink[1]{}%
\providecommand \@@endlink[0]{}%
\providecommand \url  [0]{\begingroup\@sanitize@url \@url }%
\providecommand \@url [1]{\endgroup\@href {#1}{\urlprefix }}%
\providecommand \urlprefix  [0]{URL }%
\providecommand \Eprint [0]{\href }%
\providecommand \doibase [0]{https://doi.org/}%
\providecommand \selectlanguage [0]{\@gobble}%
\providecommand \bibinfo  [0]{\@secondoftwo}%
\providecommand \bibfield  [0]{\@secondoftwo}%
\providecommand \translation [1]{[#1]}%
\providecommand \BibitemOpen [0]{}%
\providecommand \bibitemStop [0]{}%
\providecommand \bibitemNoStop [0]{.\EOS\space}%
\providecommand \EOS [0]{\spacefactor3000\relax}%
\providecommand \BibitemShut  [1]{\csname bibitem#1\endcsname}%
\let\auto@bib@innerbib\@empty
\bibitem [{\citenamefont {De~Franceschi}\ \emph {et~al.}(2010)\citenamefont
  {De~Franceschi}, \citenamefont {Kouwenhoven}, \citenamefont
  {Sch{\"o}nenberger},\ and\ \citenamefont {Wernsdorfer}}]{DeFranceschi2010}%
  \BibitemOpen
  \bibfield  {author} {\bibinfo {author} {\bibfnamefont {S.}~\bibnamefont
  {De~Franceschi}}, \bibinfo {author} {\bibfnamefont {L.~P.}\ \bibnamefont
  {Kouwenhoven}}, \bibinfo {author} {\bibfnamefont {C.}~\bibnamefont
  {Sch{\"o}nenberger}},\ and\ \bibinfo {author} {\bibfnamefont
  {W.}~\bibnamefont {Wernsdorfer}},\ }\href
  {https://doi.org/10.1038/nnano.2010.173} {\bibfield  {journal} {\bibinfo
  {journal} {Nature Nanotechnology}\ }\textbf {\bibinfo {volume} {6}},\
  \bibinfo {pages} {703} (\bibinfo {year} {2010})}\BibitemShut {NoStop}%
\bibitem [{\citenamefont {Stanescu}\ and\ \citenamefont
  {Tewari}(2013)}]{Stanescu2013}%
  \BibitemOpen
  \bibfield  {author} {\bibinfo {author} {\bibfnamefont {T.}~\bibnamefont
  {Stanescu}}\ and\ \bibinfo {author} {\bibfnamefont {S.}~\bibnamefont
  {Tewari}},\ }\href {https://doi.org/10.1088/0953-8984/25/23/233201}
  {\bibfield  {journal} {\bibinfo  {journal} {Journal of Physics-Condensed
  Matter}\ }\textbf {\bibinfo {volume} {25}},\ \bibinfo {pages} {233201}
  (\bibinfo {year} {2013})}\BibitemShut {NoStop}%
\bibitem [{\citenamefont {Lutchyn}\ \emph {et~al.}(2018)\citenamefont
  {Lutchyn}, \citenamefont {Bakkers}, \citenamefont {Kouwenhoven},
  \citenamefont {Krogstrup}, \citenamefont {Marcus},\ and\ \citenamefont
  {Oreg}}]{Lutchyn2018}%
  \BibitemOpen
  \bibfield  {author} {\bibinfo {author} {\bibfnamefont {R.}~\bibnamefont
  {Lutchyn}}, \bibinfo {author} {\bibfnamefont {E.}~\bibnamefont {Bakkers}},
  \bibinfo {author} {\bibfnamefont {L.}~\bibnamefont {Kouwenhoven}}, \bibinfo
  {author} {\bibfnamefont {P.}~\bibnamefont {Krogstrup}}, \bibinfo {author}
  {\bibnamefont {Marcus}},\ and\ \bibinfo {author} {\bibfnamefont
  {Y.}~\bibnamefont {Oreg}},\ }\href
  {https://doi.org/10.1038/s41578-018-0003-1} {\bibfield  {journal} {\bibinfo
  {journal} {Nature Review Materials}\ }\textbf {\bibinfo {volume} {3}},\
  \bibinfo {pages} {52} (\bibinfo {year} {2018})}\BibitemShut {NoStop}%
\bibitem [{\citenamefont {Sau}\ \emph {et~al.}(2010)\citenamefont {Sau},
  \citenamefont {Tewari},\ and\ \citenamefont {Das~Sarma}}]{Sau2010}%
  \BibitemOpen
  \bibfield  {author} {\bibinfo {author} {\bibfnamefont {J.~D.}\ \bibnamefont
  {Sau}}, \bibinfo {author} {\bibfnamefont {S.}~\bibnamefont {Tewari}},\ and\
  \bibinfo {author} {\bibfnamefont {S.}~\bibnamefont {Das~Sarma}},\ }\href
  {https://doi.org/10.1103/PhysRevA.82.052322} {\bibfield  {journal} {\bibinfo
  {journal} {Physical Review A}\ }\textbf {\bibinfo {volume} {82}},\ \bibinfo
  {pages} {052322} (\bibinfo {year} {2010})}\BibitemShut {NoStop}%
\bibitem [{\citenamefont {Beenakker}(2013)}]{Beenakker2013}%
  \BibitemOpen
  \bibfield  {author} {\bibinfo {author} {\bibfnamefont {C.~W.~J.}\
  \bibnamefont {Beenakker}},\ }\href
  {https://doi.org/10.1146/annurev-conmatphys-030212-184337} {\bibfield
  {journal} {\bibinfo  {journal} {Annual Review of Condensed Matter Physics}\
  }\textbf {\bibinfo {volume} {4}},\ \bibinfo {pages} {113} (\bibinfo {year}
  {2013})}\BibitemShut {NoStop}%
\bibitem [{\citenamefont {Hyart}\ \emph {et~al.}(2013)\citenamefont {Hyart},
  \citenamefont {van Heck}, \citenamefont {Fulga}, \citenamefont {Burrello},
  \citenamefont {Akhmerov},\ and\ \citenamefont {Beenakker}}]{Hyart2013}%
  \BibitemOpen
  \bibfield  {author} {\bibinfo {author} {\bibfnamefont {T.}~\bibnamefont
  {Hyart}}, \bibinfo {author} {\bibfnamefont {B.}~\bibnamefont {van Heck}},
  \bibinfo {author} {\bibfnamefont {I.~C.}\ \bibnamefont {Fulga}}, \bibinfo
  {author} {\bibfnamefont {M.}~\bibnamefont {Burrello}}, \bibinfo {author}
  {\bibfnamefont {A.~R.}\ \bibnamefont {Akhmerov}},\ and\ \bibinfo {author}
  {\bibfnamefont {C.~W.~J.}\ \bibnamefont {Beenakker}},\ }\href
  {https://doi.org/10.1103/PhysRevB.88.035121} {\bibfield  {journal} {\bibinfo
  {journal} {Physical Review B}\ }\textbf {\bibinfo {volume} {88}},\ \bibinfo
  {pages} {035121} (\bibinfo {year} {2013})}\BibitemShut {NoStop}%
\bibitem [{\citenamefont {Aasen}\ \emph {et~al.}(2016)\citenamefont {Aasen},
  \citenamefont {Hell}, \citenamefont {Mishmash}, \citenamefont {Higginbotham},
  \citenamefont {Danon}, \citenamefont {Leijnse}, \citenamefont {Jespersen},
  \citenamefont {Folk}, \citenamefont {Marcus}, \citenamefont {Flensberg},\
  and\ \citenamefont {Alicea}}]{Aasen2016}%
  \BibitemOpen
  \bibfield  {author} {\bibinfo {author} {\bibfnamefont {D.}~\bibnamefont
  {Aasen}}, \bibinfo {author} {\bibfnamefont {M.}~\bibnamefont {Hell}},
  \bibinfo {author} {\bibfnamefont {R.~V.}\ \bibnamefont {Mishmash}}, \bibinfo
  {author} {\bibfnamefont {A.}~\bibnamefont {Higginbotham}}, \bibinfo {author}
  {\bibfnamefont {J.}~\bibnamefont {Danon}}, \bibinfo {author} {\bibfnamefont
  {M.}~\bibnamefont {Leijnse}}, \bibinfo {author} {\bibfnamefont {T.~S.}\
  \bibnamefont {Jespersen}}, \bibinfo {author} {\bibfnamefont {J.~A.}\
  \bibnamefont {Folk}}, \bibinfo {author} {\bibfnamefont {C.~M.}\ \bibnamefont
  {Marcus}}, \bibinfo {author} {\bibfnamefont {K.}~\bibnamefont {Flensberg}},\
  and\ \bibinfo {author} {\bibfnamefont {J.}~\bibnamefont {Alicea}},\ }\href
  {https://doi.org/10.1103/PhysRevX.6.031016} {\bibfield  {journal} {\bibinfo
  {journal} {Physical Review X}\ }\textbf {\bibinfo {volume} {6}},\ \bibinfo
  {pages} {031016} (\bibinfo {year} {2016})}\BibitemShut {NoStop}%
\bibitem [{\citenamefont {Plugge}\ \emph {et~al.}(2017)\citenamefont {Plugge},
  \citenamefont {Rasmussen}, \citenamefont {Egger},\ and\ \citenamefont
  {Flensberg}}]{Plugge2017}%
  \BibitemOpen
  \bibfield  {author} {\bibinfo {author} {\bibfnamefont {S.}~\bibnamefont
  {Plugge}}, \bibinfo {author} {\bibfnamefont {A.}~\bibnamefont {Rasmussen}},
  \bibinfo {author} {\bibfnamefont {R.}~\bibnamefont {Egger}},\ and\ \bibinfo
  {author} {\bibfnamefont {K.}~\bibnamefont {Flensberg}},\ }\href
  {https://doi.org/10.1088/1367-2630/aa54e1} {\bibfield  {journal} {\bibinfo
  {journal} {New Journal of Physics}\ }\textbf {\bibinfo {volume} {19}},\
  \bibinfo {pages} {012001} (\bibinfo {year} {2017})}\BibitemShut {NoStop}%
\bibitem [{\citenamefont {Leijnse}\ and\ \citenamefont
  {Flensberg}(2011)}]{Leijnse2011}%
  \BibitemOpen
  \bibfield  {author} {\bibinfo {author} {\bibfnamefont {M.}~\bibnamefont
  {Leijnse}}\ and\ \bibinfo {author} {\bibfnamefont {K.}~\bibnamefont
  {Flensberg}},\ }\href {https://doi.org/10.1103/PhysRevB.84.140501} {\bibfield
   {journal} {\bibinfo  {journal} {Physical Review B}\ }\textbf {\bibinfo
  {volume} {84}},\ \bibinfo {pages} {140501(R)} (\bibinfo {year}
  {2011})}\BibitemShut {NoStop}%
\bibitem [{\citenamefont {Liu}\ and\ \citenamefont {Baranger}(2011)}]{Liu2011}%
  \BibitemOpen
  \bibfield  {author} {\bibinfo {author} {\bibfnamefont {D.~E.}\ \bibnamefont
  {Liu}}\ and\ \bibinfo {author} {\bibfnamefont {H.~U.}\ \bibnamefont
  {Baranger}},\ }\href {https://doi.org/10.1103/PhysRevB.84.201308} {\bibfield
  {journal} {\bibinfo  {journal} {Physical Review B}\ }\textbf {\bibinfo
  {volume} {84}},\ \bibinfo {pages} {201308(R)} (\bibinfo {year}
  {2011})}\BibitemShut {NoStop}%
\bibitem [{\citenamefont {Deng}\ \emph {et~al.}(2016)\citenamefont {Deng},
  \citenamefont {Vaitiekėnas}, \citenamefont {Hansen}, \citenamefont {Danon},
  \citenamefont {Leijnse}, \citenamefont {Flensberg}, \citenamefont {Nygard},
  \citenamefont {Krogstrup},\ and\ \citenamefont {Marcus}}]{Deng2016}%
  \BibitemOpen
  \bibfield  {author} {\bibinfo {author} {\bibfnamefont {M.~T.}\ \bibnamefont
  {Deng}}, \bibinfo {author} {\bibfnamefont {S.}~\bibnamefont {Vaitiekėnas}},
  \bibinfo {author} {\bibfnamefont {E.~B.}\ \bibnamefont {Hansen}}, \bibinfo
  {author} {\bibfnamefont {J.}~\bibnamefont {Danon}}, \bibinfo {author}
  {\bibfnamefont {M.}~\bibnamefont {Leijnse}}, \bibinfo {author} {\bibfnamefont
  {K.}~\bibnamefont {Flensberg}}, \bibinfo {author} {\bibfnamefont
  {J.}~\bibnamefont {Nygard}}, \bibinfo {author} {\bibfnamefont
  {P.}~\bibnamefont {Krogstrup}},\ and\ \bibinfo {author} {\bibfnamefont
  {C.~M.}\ \bibnamefont {Marcus}},\ }\href
  {https://doi.org/10.1126/science.aaf3961} {\bibfield  {journal} {\bibinfo
  {journal} {Science}\ }\textbf {\bibinfo {volume} {23}},\ \bibinfo {pages}
  {1557} (\bibinfo {year} {2016})}\BibitemShut {NoStop}%
\bibitem [{\citenamefont {Liu}\ \emph {et~al.}(2017)\citenamefont {Liu},
  \citenamefont {Sau}, \citenamefont {Stanescu},\ and\ \citenamefont
  {Das~Sarma}}]{Liu2016}%
  \BibitemOpen
  \bibfield  {author} {\bibinfo {author} {\bibfnamefont {C.~X.}\ \bibnamefont
  {Liu}}, \bibinfo {author} {\bibfnamefont {J.~D.}\ \bibnamefont {Sau}},
  \bibinfo {author} {\bibfnamefont {T.~D.}\ \bibnamefont {Stanescu}},\ and\
  \bibinfo {author} {\bibfnamefont {S.}~\bibnamefont {Das~Sarma}},\ }\href
  {https://doi.org/10.1103/PhysRevB.96.075161} {\bibfield  {journal} {\bibinfo
  {journal} {Physical Review B}\ }\textbf {\bibinfo {volume} {96}},\ \bibinfo
  {pages} {075161} (\bibinfo {year} {2017})}\BibitemShut {NoStop}%
\bibitem [{\citenamefont {Seifert}\ \emph {et~al.}(2004)\citenamefont
  {Seifert}, \citenamefont {Borgstrom}, \citenamefont {Deppert}, \citenamefont
  {A.}, \citenamefont {Johansson}, \citenamefont {W.}, \citenamefont
  {Martensson}, \citenamefont {Skold}, \citenamefont {Svensson}, \citenamefont
  {Wacaser}, \citenamefont {Wallenberg},\ and\ \citenamefont
  {Samuelson}}]{Seifert2004}%
  \BibitemOpen
  \bibfield  {author} {\bibinfo {author} {\bibfnamefont {W.}~\bibnamefont
  {Seifert}}, \bibinfo {author} {\bibfnamefont {M.}~\bibnamefont {Borgstrom}},
  \bibinfo {author} {\bibfnamefont {K.}~\bibnamefont {Deppert}}, \bibinfo
  {author} {\bibfnamefont {D.~K.}\ \bibnamefont {A.}}, \bibinfo {author}
  {\bibfnamefont {J.}~\bibnamefont {Johansson}}, \bibinfo {author}
  {\bibfnamefont {L.~M.}\ \bibnamefont {W.}}, \bibinfo {author} {\bibfnamefont
  {T.}~\bibnamefont {Martensson}}, \bibinfo {author} {\bibfnamefont
  {N.}~\bibnamefont {Skold}}, \bibinfo {author} {\bibfnamefont {C.~P.~T.}\
  \bibnamefont {Svensson}}, \bibinfo {author} {\bibfnamefont {B.~A.}\
  \bibnamefont {Wacaser}}, \bibinfo {author} {\bibfnamefont {L.~R.}\
  \bibnamefont {Wallenberg}},\ and\ \bibinfo {author} {\bibfnamefont
  {L.}~\bibnamefont {Samuelson}},\ }\href
  {https://doi.org/10.1016/j.jcrysgro.2004.09.023} {\bibfield  {journal}
  {\bibinfo  {journal} {Journal of Crystal Growth}\ }\textbf {\bibinfo {volume}
  {272}},\ \bibinfo {pages} {211} (\bibinfo {year} {2004})}\BibitemShut
  {NoStop}%
\bibitem [{\citenamefont {Björk}\ \emph {et~al.}(2002)\citenamefont {Björk},
  \citenamefont {Ohlsson}, \citenamefont {Thelander}, \citenamefont {Persson},
  \citenamefont {Deppert}, \citenamefont {Wallenberg},\ and\ \citenamefont
  {Samuelson}}]{Bjork2002}%
  \BibitemOpen
  \bibfield  {author} {\bibinfo {author} {\bibfnamefont {M.~T.}\ \bibnamefont
  {Björk}}, \bibinfo {author} {\bibfnamefont {B.~J.}\ \bibnamefont {Ohlsson}},
  \bibinfo {author} {\bibfnamefont {C.}~\bibnamefont {Thelander}}, \bibinfo
  {author} {\bibfnamefont {A.~I.}\ \bibnamefont {Persson}}, \bibinfo {author}
  {\bibfnamefont {K.}~\bibnamefont {Deppert}}, \bibinfo {author} {\bibfnamefont
  {L.~R.}\ \bibnamefont {Wallenberg}},\ and\ \bibinfo {author} {\bibfnamefont
  {L.}~\bibnamefont {Samuelson}},\ }\href {https://doi.org/10.1063/1.1527995}
  {\bibfield  {journal} {\bibinfo  {journal} {Applied Physics Letters}\
  }\textbf {\bibinfo {volume} {81}},\ \bibinfo {pages} {4458} (\bibinfo {year}
  {2002})}\BibitemShut {NoStop}%
\bibitem [{\citenamefont {Niquet}\ and\ \citenamefont
  {Mojica}(2008)}]{Niquet2008}%
  \BibitemOpen
  \bibfield  {author} {\bibinfo {author} {\bibfnamefont {Y.-M.}\ \bibnamefont
  {Niquet}}\ and\ \bibinfo {author} {\bibfnamefont {D.~C.}\ \bibnamefont
  {Mojica}},\ }\href {https://doi.org/10.1103/PhysRevB.77.115316} {\bibfield
  {journal} {\bibinfo  {journal} {Physical Review B}\ }\textbf {\bibinfo
  {volume} {77}},\ \bibinfo {pages} {115316} (\bibinfo {year}
  {2008})}\BibitemShut {NoStop}%
\bibitem [{\citenamefont {Thomas}\ \emph {et~al.}(2020)\citenamefont {Thomas},
  \citenamefont {Baumgartner}, \citenamefont {Gubser}, \citenamefont
  {J{\"u}nger}, \citenamefont {F{\"u}l{\"o}p}, \citenamefont {Nilsson},
  \citenamefont {Rossi}, \citenamefont {Zannier}, \citenamefont {Sorba},\ and\
  \citenamefont {Sch{\"o}nenberger}}]{Thomas2020}%
  \BibitemOpen
  \bibfield  {author} {\bibinfo {author} {\bibfnamefont {F.~S.}\ \bibnamefont
  {Thomas}}, \bibinfo {author} {\bibfnamefont {A.}~\bibnamefont {Baumgartner}},
  \bibinfo {author} {\bibfnamefont {L.}~\bibnamefont {Gubser}}, \bibinfo
  {author} {\bibfnamefont {C.}~\bibnamefont {J{\"u}nger}}, \bibinfo {author}
  {\bibfnamefont {G.}~\bibnamefont {F{\"u}l{\"o}p}}, \bibinfo {author}
  {\bibfnamefont {N.}~\bibnamefont {Nilsson}}, \bibinfo {author} {\bibfnamefont
  {F.}~\bibnamefont {Rossi}}, \bibinfo {author} {\bibfnamefont
  {V.}~\bibnamefont {Zannier}}, \bibinfo {author} {\bibfnamefont
  {L.}~\bibnamefont {Sorba}},\ and\ \bibinfo {author} {\bibfnamefont
  {C.}~\bibnamefont {Sch{\"o}nenberger}},\ }\href
  {https://doi.org/10.1088/1361-6528/ab5ce6} {\bibfield  {journal} {\bibinfo
  {journal} {Nanotechnology}\ }\textbf {\bibinfo {volume} {31}},\ \bibinfo
  {pages} {135003} (\bibinfo {year} {2020})}\BibitemShut {NoStop}%
\bibitem [{\citenamefont {Nilsson}\ \emph {et~al.}(2016)\citenamefont
  {Nilsson}, \citenamefont {Namazi}, \citenamefont {Lehmann}, \citenamefont
  {Leijnse}, \citenamefont {Dick},\ and\ \citenamefont
  {Thelander}}]{Nilsson2016}%
  \BibitemOpen
  \bibfield  {author} {\bibinfo {author} {\bibfnamefont {M.}~\bibnamefont
  {Nilsson}}, \bibinfo {author} {\bibfnamefont {L.}~\bibnamefont {Namazi}},
  \bibinfo {author} {\bibfnamefont {S.}~\bibnamefont {Lehmann}}, \bibinfo
  {author} {\bibfnamefont {M.}~\bibnamefont {Leijnse}}, \bibinfo {author}
  {\bibfnamefont {K.~A.}\ \bibnamefont {Dick}},\ and\ \bibinfo {author}
  {\bibfnamefont {C.}~\bibnamefont {Thelander}},\ }\href
  {https://doi.org/10.1103/PhysRevB.93.195422} {\bibfield  {journal} {\bibinfo
  {journal} {Phys. Rev. B}\ }\textbf {\bibinfo {volume} {93}},\ \bibinfo
  {pages} {195422} (\bibinfo {year} {2016})}\BibitemShut {NoStop}%
\bibitem [{\citenamefont {J{\"u}nger}\ \emph {et~al.}(2019)\citenamefont
  {J{\"u}nger}, \citenamefont {Baumgartner}, \citenamefont {Delagrange},
  \citenamefont {Chevallier}, \citenamefont {Lehmann}, \citenamefont {Nilsson},
  \citenamefont {Dick}, \citenamefont {Thelander},\ and\ \citenamefont
  {Sch{\"o}nenberger}}]{Junger2019}%
  \BibitemOpen
  \bibfield  {author} {\bibinfo {author} {\bibfnamefont {C.}~\bibnamefont
  {J{\"u}nger}}, \bibinfo {author} {\bibfnamefont {A.}~\bibnamefont
  {Baumgartner}}, \bibinfo {author} {\bibfnamefont {R.}~\bibnamefont
  {Delagrange}}, \bibinfo {author} {\bibfnamefont {D.}~\bibnamefont
  {Chevallier}}, \bibinfo {author} {\bibfnamefont {S.}~\bibnamefont {Lehmann}},
  \bibinfo {author} {\bibfnamefont {M.}~\bibnamefont {Nilsson}}, \bibinfo
  {author} {\bibfnamefont {K.~A.}\ \bibnamefont {Dick}}, \bibinfo {author}
  {\bibfnamefont {C.}~\bibnamefont {Thelander}},\ and\ \bibinfo {author}
  {\bibfnamefont {C.}~\bibnamefont {Sch{\"o}nenberger}},\ }\href
  {https://doi.org/10.1038/s42005-019-0162-4} {\bibfield  {journal} {\bibinfo
  {journal} {Communications Physics}\ }\textbf {\bibinfo {volume} {2}}
  (\bibinfo {year} {2019})}\BibitemShut {NoStop}%
\bibitem [{\citenamefont {J{\"u}nger}\ \emph {et~al.}(2020)\citenamefont
  {J{\"u}nger}, \citenamefont {Delagrange}, \citenamefont {Chevallier},
  \citenamefont {Lehmann}, \citenamefont {Dick}, \citenamefont {Thelander},
  \citenamefont {Klinovaja}, \citenamefont {Loss}, \citenamefont
  {Baumgartner},\ and\ \citenamefont {Sch{\"o}nenberger}}]{Junger2020}%
  \BibitemOpen
  \bibfield  {author} {\bibinfo {author} {\bibfnamefont {C.}~\bibnamefont
  {J{\"u}nger}}, \bibinfo {author} {\bibfnamefont {R.}~\bibnamefont
  {Delagrange}}, \bibinfo {author} {\bibfnamefont {D.}~\bibnamefont
  {Chevallier}}, \bibinfo {author} {\bibfnamefont {S.}~\bibnamefont {Lehmann}},
  \bibinfo {author} {\bibfnamefont {K.~A.}\ \bibnamefont {Dick}}, \bibinfo
  {author} {\bibfnamefont {C.}~\bibnamefont {Thelander}}, \bibinfo {author}
  {\bibfnamefont {J.}~\bibnamefont {Klinovaja}}, \bibinfo {author}
  {\bibfnamefont {D.}~\bibnamefont {Loss}}, \bibinfo {author} {\bibfnamefont
  {A.}~\bibnamefont {Baumgartner}},\ and\ \bibinfo {author} {\bibfnamefont
  {C.}~\bibnamefont {Sch{\"o}nenberger}},\ }\href
  {https://doi.org/10.1103/PhysRevLett.125.017701} {\bibfield  {journal}
  {\bibinfo  {journal} {Physical Review Letters}\ }\textbf {\bibinfo {volume}
  {125}},\ \bibinfo {pages} {017701} (\bibinfo {year} {2020})}\BibitemShut
  {NoStop}%
\bibitem [{\citenamefont {Yu}(1965)}]{Yu1965}%
  \BibitemOpen
  \bibfield  {author} {\bibinfo {author} {\bibfnamefont {L.}~\bibnamefont
  {Yu}},\ }\href {http://wulixb.iphy.ac.cn/CN/Y1965/V21/I1/75} {\bibfield
  {journal} {\bibinfo  {journal} {Acta Physica Sinica}\ }\textbf {\bibinfo
  {volume} {21}},\ \bibinfo {pages} {57} (\bibinfo {year} {1965})}\BibitemShut
  {NoStop}%
\bibitem [{\citenamefont {Shiba}(1968)}]{Shiba1968}%
  \BibitemOpen
  \bibfield  {author} {\bibinfo {author} {\bibfnamefont {H.}~\bibnamefont
  {Shiba}},\ }\href {https://doi.org/10.1143/PTP.40.435} {\bibfield  {journal}
  {\bibinfo  {journal} {Progress of Theoretical Physics}\ }\textbf {\bibinfo
  {volume} {40}},\ \bibinfo {pages} {453} (\bibinfo {year} {1968})}\BibitemShut
  {NoStop}%
\bibitem [{\citenamefont {Rusinov}(1969)}]{Rusinov1969}%
  \BibitemOpen
  \bibfield  {author} {\bibinfo {author} {\bibfnamefont {A.~I.}\ \bibnamefont
  {Rusinov}},\ }\href {http://www.jetp.ac.ru/cgi-bin/dn/e_029_06_1101.pdf}
  {\bibfield  {journal} {\bibinfo  {journal} {Sov. J. Exp. Theor. Phys.}\
  }\textbf {\bibinfo {volume} {29}},\ \bibinfo {pages} {1101} (\bibinfo {year}
  {1969})}\BibitemShut {NoStop}%
\bibitem [{\citenamefont {Thelander}\ \emph {et~al.}(2003)\citenamefont
  {Thelander}, \citenamefont {Bj{\"o}rk}, \citenamefont {Ohlsson},
  \citenamefont {Larsson}, \citenamefont {Wallenberg},\ and\ \citenamefont
  {Samuelson}}]{Bjork2003}%
  \BibitemOpen
  \bibfield  {author} {\bibinfo {author} {\bibfnamefont {T.}~\bibnamefont
  {Thelander}, \bibfnamefont {C.~T.and~M{\aa}rtensson}}, \bibinfo {author}
  {\bibfnamefont {M.~T.}\ \bibnamefont {Bj{\"o}rk}}, \bibinfo {author}
  {\bibfnamefont {B.~J.}\ \bibnamefont {Ohlsson}}, \bibinfo {author}
  {\bibfnamefont {M.~W.}\ \bibnamefont {Larsson}}, \bibinfo {author}
  {\bibfnamefont {L.~R.}\ \bibnamefont {Wallenberg}},\ and\ \bibinfo {author}
  {\bibfnamefont {L.}~\bibnamefont {Samuelson}},\ }\href
  {https://doi.org/10.1063/1.1606889} {\bibfield  {journal} {\bibinfo
  {journal} {Applied Physics Letters}\ }\textbf {\bibinfo {volume} {83}},\
  \bibinfo {pages} {2052} (\bibinfo {year} {2003})}\BibitemShut {NoStop}%
\bibitem [{\citenamefont {Bj{\"o}rk}\ \emph {et~al.}(2004)\citenamefont
  {Bj{\"o}rk}, \citenamefont {Thelander}, \citenamefont {Hansen}, \citenamefont
  {Jensen}, \citenamefont {Larsson}, \citenamefont {Wallenberg},\ and\
  \citenamefont {Samuelson}}]{Bjork2004}%
  \BibitemOpen
  \bibfield  {author} {\bibinfo {author} {\bibfnamefont {M.~T.}\ \bibnamefont
  {Bj{\"o}rk}}, \bibinfo {author} {\bibfnamefont {C.}~\bibnamefont
  {Thelander}}, \bibinfo {author} {\bibfnamefont {A.~E.}\ \bibnamefont
  {Hansen}}, \bibinfo {author} {\bibfnamefont {L.~E.}\ \bibnamefont {Jensen}},
  \bibinfo {author} {\bibnamefont {Larsson}}, \bibinfo {author} {\bibfnamefont
  {M.~W.}\ \bibnamefont {Wallenberg}},\ and\ \bibinfo {author} {\bibfnamefont
  {L.}~\bibnamefont {Samuelson}},\ }\href {https://doi.org/10.1021/nl049230s}
  {\bibfield  {journal} {\bibinfo  {journal} {Nano Letters}\ }\textbf {\bibinfo
  {volume} {4}},\ \bibinfo {pages} {1621} (\bibinfo {year} {2004})}\BibitemShut
  {NoStop}%
\bibitem [{\citenamefont {Bj{\"o}rk}\ \emph {et~al.}(2005)\citenamefont
  {Bj{\"o}rk}, \citenamefont {Fuhrer}, \citenamefont {Hansen}, \citenamefont
  {Larsson}, \citenamefont {Fr\"oberg},\ and\ \citenamefont
  {Samuelson}}]{Bjork2005}%
  \BibitemOpen
  \bibfield  {author} {\bibinfo {author} {\bibfnamefont {M.~T.}\ \bibnamefont
  {Bj{\"o}rk}}, \bibinfo {author} {\bibfnamefont {A.}~\bibnamefont {Fuhrer}},
  \bibinfo {author} {\bibfnamefont {A.~E.}\ \bibnamefont {Hansen}}, \bibinfo
  {author} {\bibfnamefont {M.~W.}\ \bibnamefont {Larsson}}, \bibinfo {author}
  {\bibfnamefont {L.~E.}\ \bibnamefont {Fr\"oberg}},\ and\ \bibinfo {author}
  {\bibfnamefont {L.}~\bibnamefont {Samuelson}},\ }\href
  {https://doi.org/10.1103/PhysRevB.72.201307} {\bibfield  {journal} {\bibinfo
  {journal} {Phys. Rev. B}\ }\textbf {\bibinfo {volume} {72}},\ \bibinfo
  {pages} {201307(R)} (\bibinfo {year} {2005})}\BibitemShut {NoStop}%
\bibitem [{\citenamefont {Momtaz}\ \emph {et~al.}(2020)\citenamefont {Momtaz},
  \citenamefont {Servino}, \citenamefont {Demontis}, \citenamefont {Zannier},
  \citenamefont {Ercolani}, \citenamefont {Rossi}, \citenamefont {Rossella},
  \citenamefont {Sorba}, \citenamefont {Beltram},\ and\ \citenamefont
  {Roddaro}}]{Momtaz2020}%
  \BibitemOpen
  \bibfield  {author} {\bibinfo {author} {\bibfnamefont {Z.~S.}\ \bibnamefont
  {Momtaz}}, \bibinfo {author} {\bibfnamefont {S.}~\bibnamefont {Servino}},
  \bibinfo {author} {\bibfnamefont {V.}~\bibnamefont {Demontis}}, \bibinfo
  {author} {\bibfnamefont {V.}~\bibnamefont {Zannier}}, \bibinfo {author}
  {\bibfnamefont {D.}~\bibnamefont {Ercolani}}, \bibinfo {author}
  {\bibfnamefont {F.}~\bibnamefont {Rossi}}, \bibinfo {author} {\bibfnamefont
  {F.}~\bibnamefont {Rossella}}, \bibinfo {author} {\bibfnamefont
  {L.}~\bibnamefont {Sorba}}, \bibinfo {author} {\bibfnamefont
  {F.}~\bibnamefont {Beltram}},\ and\ \bibinfo {author} {\bibfnamefont
  {S.}~\bibnamefont {Roddaro}},\ }\href
  {https://doi.org/10.1021/acs.nanolett.9b04850} {\bibfield  {journal}
  {\bibinfo  {journal} {Nano Letters}\ }\textbf {\bibinfo {volume} {20}},\
  \bibinfo {pages} {1693} (\bibinfo {year} {2020})}\BibitemShut {NoStop}%
\bibitem [{\citenamefont {M\"ott\"onen}\ \emph {et~al.}(2010)\citenamefont
  {M\"ott\"onen}, \citenamefont {Tan}, \citenamefont {Chan}, \citenamefont
  {Zwanenburg}, \citenamefont {Lim}, \citenamefont {Escott}, \citenamefont
  {Pirkkalainen}, \citenamefont {Morello}, \citenamefont {Yang}, \citenamefont
  {van Donkelaar}, \citenamefont {Alves}, \citenamefont {Jamieson},
  \citenamefont {Hollenberg},\ and\ \citenamefont {Dzurak}}]{Mottonen2010}%
  \BibitemOpen
  \bibfield  {author} {\bibinfo {author} {\bibfnamefont {M.}~\bibnamefont
  {M\"ott\"onen}}, \bibinfo {author} {\bibfnamefont {K.~Y.}\ \bibnamefont
  {Tan}}, \bibinfo {author} {\bibfnamefont {K.~W.}\ \bibnamefont {Chan}},
  \bibinfo {author} {\bibfnamefont {F.~A.}\ \bibnamefont {Zwanenburg}},
  \bibinfo {author} {\bibfnamefont {W.~H.}\ \bibnamefont {Lim}}, \bibinfo
  {author} {\bibfnamefont {C.~C.}\ \bibnamefont {Escott}}, \bibinfo {author}
  {\bibfnamefont {J.-M.}\ \bibnamefont {Pirkkalainen}}, \bibinfo {author}
  {\bibfnamefont {A.}~\bibnamefont {Morello}}, \bibinfo {author} {\bibfnamefont
  {C.}~\bibnamefont {Yang}}, \bibinfo {author} {\bibfnamefont {J.~A.}\
  \bibnamefont {van Donkelaar}}, \bibinfo {author} {\bibfnamefont {A.~D.~C.}\
  \bibnamefont {Alves}}, \bibinfo {author} {\bibfnamefont {D.~N.}\ \bibnamefont
  {Jamieson}}, \bibinfo {author} {\bibfnamefont {L.~C.~L.}\ \bibnamefont
  {Hollenberg}},\ and\ \bibinfo {author} {\bibfnamefont {A.~S.}\ \bibnamefont
  {Dzurak}},\ }\href {https://doi.org/10.1103/PhysRevB.81.161304} {\bibfield
  {journal} {\bibinfo  {journal} {Phys. Rev. B}\ }\textbf {\bibinfo {volume}
  {81}},\ \bibinfo {pages} {161304(R)} (\bibinfo {year} {2010})}\BibitemShut
  {NoStop}%
\bibitem [{\citenamefont {Gehring}\ \emph {et~al.}(2017)\citenamefont
  {Gehring}, \citenamefont {Sowa}, \citenamefont {Cremers}, \citenamefont {Wu},
  \citenamefont {Sadeghi}, \citenamefont {Sheng}, \citenamefont {Warner},
  \citenamefont {Lambert}, \citenamefont {Briggs},\ and\ \citenamefont
  {Mol}}]{Gehring2017}%
  \BibitemOpen
  \bibfield  {author} {\bibinfo {author} {\bibfnamefont {P.}~\bibnamefont
  {Gehring}}, \bibinfo {author} {\bibfnamefont {J.~K.}\ \bibnamefont {Sowa}},
  \bibinfo {author} {\bibfnamefont {J.}~\bibnamefont {Cremers}}, \bibinfo
  {author} {\bibfnamefont {Q.}~\bibnamefont {Wu}}, \bibinfo {author}
  {\bibfnamefont {H.}~\bibnamefont {Sadeghi}}, \bibinfo {author} {\bibfnamefont
  {Y.}~\bibnamefont {Sheng}}, \bibinfo {author} {\bibfnamefont {J.~H.}\
  \bibnamefont {Warner}}, \bibinfo {author} {\bibfnamefont {C.~J.}\
  \bibnamefont {Lambert}}, \bibinfo {author} {\bibfnamefont {G.~A.~D.}\
  \bibnamefont {Briggs}},\ and\ \bibinfo {author} {\bibfnamefont {J.~A.}\
  \bibnamefont {Mol}},\ }\href {https://doi.org/10.1021/acsnano.7b00570}
  {\bibfield  {journal} {\bibinfo  {journal} {ACS Nano}\ }\textbf {\bibinfo
  {volume} {11}},\ \bibinfo {pages} {5325–5331} (\bibinfo {year}
  {2017})}\BibitemShut {NoStop}%
\bibitem [{\citenamefont {Zannier}\ \emph {et~al.}(2019)\citenamefont
  {Zannier}, \citenamefont {Rossi}, \citenamefont {Ercolani},\ and\
  \citenamefont {Sorba}}]{Zannier2019}%
  \BibitemOpen
  \bibfield  {author} {\bibinfo {author} {\bibfnamefont {V.}~\bibnamefont
  {Zannier}}, \bibinfo {author} {\bibfnamefont {F.}~\bibnamefont {Rossi}},
  \bibinfo {author} {\bibfnamefont {D.}~\bibnamefont {Ercolani}},\ and\
  \bibinfo {author} {\bibfnamefont {L.}~\bibnamefont {Sorba}},\ }\href
  {https://doi.org/10.1088/1361-6528/aaf7ab} {\bibfield  {journal} {\bibinfo
  {journal} {Nanotechnology}\ }\textbf {\bibinfo {volume} {30}},\ \bibinfo
  {pages} {094003} (\bibinfo {year} {2019})}\BibitemShut {NoStop}%
\bibitem [{\citenamefont {Suyatin}\ \emph {et~al.}(2007)\citenamefont
  {Suyatin}, \citenamefont {Thelander}, \citenamefont {Bj{\"o}rk},
  \citenamefont {Maximov},\ and\ \citenamefont {Samuelson}}]{Suyatin2007}%
  \BibitemOpen
  \bibfield  {author} {\bibinfo {author} {\bibfnamefont {D.~B.}\ \bibnamefont
  {Suyatin}}, \bibinfo {author} {\bibfnamefont {C.}~\bibnamefont {Thelander}},
  \bibinfo {author} {\bibfnamefont {M.~T.}\ \bibnamefont {Bj{\"o}rk}}, \bibinfo
  {author} {\bibfnamefont {I.}~\bibnamefont {Maximov}},\ and\ \bibinfo {author}
  {\bibfnamefont {L.}~\bibnamefont {Samuelson}},\ }\href
  {https://doi.org/10.1088/0957-4484/18/10/105307} {\bibfield  {journal}
  {\bibinfo  {journal} {Nanotechnology}\ }\textbf {\bibinfo {volume} {18}},\
  \bibinfo {pages} {105307} (\bibinfo {year} {2007})}\BibitemShut {NoStop}%
\bibitem [{\citenamefont {Court}\ \emph {et~al.}(2007)\citenamefont {Court},
  \citenamefont {Ferguson},\ and\ \citenamefont {Clark}}]{Court2007}%
  \BibitemOpen
  \bibfield  {author} {\bibinfo {author} {\bibfnamefont {N.~A.}\ \bibnamefont
  {Court}}, \bibinfo {author} {\bibfnamefont {A.~J.}\ \bibnamefont
  {Ferguson}},\ and\ \bibinfo {author} {\bibfnamefont {R.~G.}\ \bibnamefont
  {Clark}},\ }\href {https://doi.org/10.1088/0953-2048/21/01/015013} {\bibfield
   {journal} {\bibinfo  {journal} {Superconductor Science and Technology}\
  }\textbf {\bibinfo {volume} {21}},\ \bibinfo {pages} {015013} (\bibinfo
  {year} {2007})}\BibitemShut {NoStop}%
\bibitem [{\citenamefont {Kouwenhoven}\ \emph {et~al.}(1997)\citenamefont
  {Kouwenhoven}, \citenamefont {Marcus}, \citenamefont {McEuen}, ,
  \citenamefont {Tarucha}, \citenamefont {Westervelt},\ and\ \citenamefont
  {Wingreen}}]{Kouwenhoven1997}%
  \BibitemOpen
  \bibfield  {author} {\bibinfo {author} {\bibfnamefont {L.~P.}\ \bibnamefont
  {Kouwenhoven}}, \bibinfo {author} {\bibfnamefont {C.~M.}\ \bibnamefont
  {Marcus}}, \bibinfo {author} {\bibfnamefont {P.~L.}\ \bibnamefont {McEuen}},
  , \bibinfo {author} {\bibfnamefont {S.}~\bibnamefont {Tarucha}}, \bibinfo
  {author} {\bibfnamefont {R.~M.}\ \bibnamefont {Westervelt}},\ and\ \bibinfo
  {author} {\bibfnamefont {N.~S.}\ \bibnamefont {Wingreen}},\ }\bibinfo {title}
  {Electron transport in quantum dots},\ in\ \href
  {https://doi.org/10.1007/978-94-015-8839-3_4} {\emph {\bibinfo {booktitle}
  {Mesoscopic Electron Transport}}}\ (\bibinfo  {publisher} {Springer
  Netherlands},\ \bibinfo {address} {Dordrecht},\ \bibinfo {year} {1997})\ pp.\
  \bibinfo {pages} {105--214}\BibitemShut {NoStop}%
\bibitem [{\citenamefont {Beenakker}(1991)}]{Beenakker1991}%
  \BibitemOpen
  \bibfield  {author} {\bibinfo {author} {\bibfnamefont {C.~W.~J.}\
  \bibnamefont {Beenakker}},\ }\href {https://doi.org/10.1103/PhysRevB.44.1646}
  {\bibfield  {journal} {\bibinfo  {journal} {Phys. Rev. B}\ }\textbf {\bibinfo
  {volume} {44}},\ \bibinfo {pages} {1646} (\bibinfo {year}
  {1991})}\BibitemShut {NoStop}%
\bibitem [{\citenamefont {Bar-Ad}\ \emph {et~al.}(1997)\citenamefont {Bar-Ad},
  \citenamefont {Kner}, \citenamefont {Marquezini}, \citenamefont {Mukamel},\
  and\ \citenamefont {Chemla}}]{Bar-Ad1997}%
  \BibitemOpen
  \bibfield  {author} {\bibinfo {author} {\bibfnamefont {S.}~\bibnamefont
  {Bar-Ad}}, \bibinfo {author} {\bibfnamefont {P.}~\bibnamefont {Kner}},
  \bibinfo {author} {\bibfnamefont {M.~V.}\ \bibnamefont {Marquezini}},
  \bibinfo {author} {\bibfnamefont {S.}~\bibnamefont {Mukamel}},\ and\ \bibinfo
  {author} {\bibfnamefont {D.~S.}\ \bibnamefont {Chemla}},\ }\href
  {https://doi.org/10.1103/PhysRevLett.78.1363} {\bibfield  {journal} {\bibinfo
   {journal} {Phys. Rev. Lett.}\ }\textbf {\bibinfo {volume} {78}},\ \bibinfo
  {pages} {1363} (\bibinfo {year} {1997})}\BibitemShut {NoStop}%
\end{thebibliography}

%

\end{document}